\documentclass[12pt]{article}

\usepackage{scicite}
\usepackage{times}

\usepackage{aas_macros}
\usepackage{amssymb}
\usepackage{amsmath}
\usepackage{wasysym}
\usepackage{graphicx}
\usepackage{xspace}
\usepackage[usenames, dvipsnames]{color}
\usepackage[normalem]{ulem}
\usepackage{soul}
\usepackage{nicefrac}
\usepackage{makecell}
\usepackage{url}
\usepackage{xr-hyper}
\usepackage{hyperref}
\usepackage{footmisc}
\usepackage{pgffor}
\usepackage{xcolor}
\usepackage{subfiles} % to enable independent compilation of the main article and som 
\usepackage{subfig}
\usepackage{setspace}
\usepackage[font=footnotesize]{caption}
\makeatletter
\newcommand*{\addFileDependency}[1]{% argument=file name and extension
  \typeout{(#1)}
  \@addtofilelist{#1}
  \IfFileExists{#1}{}{\typeout{No file #1.}}
}
\def\myd{\mathrm{d}}
\def\myd{{\rm d}}
\def\dif{\@ifnextchar[{\@with}{\@without}}

\def\@with[#1]#2{
  \ensuremath{
    \mathchoice
    {\frac{\foreach \x in {#2}{\,\myd\x}}{\foreach \x in {#1}{\myd\x\,}}}%
    {{\foreach \x in {#2}{\,\myd\x}}/{\foreach \x in {#1}{\,\myd\x}}}%
    {{\foreach \x in {#2}{\,\myd\x}}/{\foreach \x in {#1}{\,\myd\x}}}%
    {{\foreach \x in {#2}{\,\myd\x}}/{\foreach \x in {#1}{\,\myd\x}}}
  }
}

\def\@without#1{
  \ensuremath{%
    \ifx\hfuzz#1\hfuzz
    \myd
    \else
    \foreach \x in {#1}{\,\myd\x}
    \fi
    }
}

\makeatother

\usepackage{booktabs,multirow}
\usepackage{floatrow}
\DeclareFloatFont{tiny}{\scriptsize}
\newenvironment{sciabstract}{%
\begin{quote} \bf}
{\end{quote}}

\title
{Time-resolved hadronic particle acceleration in the recurrent nova RS\,Ophiuchi}

\author{H.E.S.S. Collaboration\footnote{alison.mw.mitchell@fau.de, stefan.ohm@desy.de, brian.reville@mpi-hd.mpg.de,
\newline
\indent\,\,\, contact.hess@hess-experiment.eu}
\footnote{H.E.S.S. Collaboration authors and affiliations are listed in the supplementary materials}}

\date{}

\begin{document}

\baselineskip24pt

\maketitle

\begin{sciabstract}
Recurrent novae are repeating thermonuclear explosions in the outer layers of white dwarfs, due to the accretion of fresh material from a binary companion. The shock generated by ejected material slamming into the companion star's wind, accelerates particles to very-high-energies. We report very-high-energy (VHE, $\gtrsim100$\,GeV) gamma rays from the recurrent nova RS\,Ophiuchi up to a month after its 2021 outburst, using the High Energy Stereoscopic System.
The VHE emission has a similar temporal profile to lower-energy GeV emission, indicating a common origin, with a two-day delay in peak flux. 
These observations constrain models of time-dependent particle energization, favouring a hadronic emission scenario over the leptonic alternative.
This confirms that shocks in dense winds provide favourable environments for efficient cosmic-ray acceleration to very-high-energies. 
\end{sciabstract}

\newpage 

%Introduction
RS Ophiuchi (RS Oph) is a recurrent nova system
comprising a white dwarf and a companion red giant star. Novae are a source of high-energy particles \cite{2010Sci...329..817A,2014Sci...345..554A}, with non-thermal gamma-ray emission in the range $\sim$100\,MeV to $\sim$10\,GeV \cite{ChomiukReview}.
The RS Oph system is
located approximately $1.4$kpc from Earth \cite{2008ASPC..401...52B}; analysis of Gaia data suggests larger distances of $2.3$ kpc \cite{2018A&A...616A...1G} or $2.7$ kpc \cite{2021A&A...649A...4L}, although the reliability of these estimates is questionable due to the orbital motion in RS Oph. 
The binary components have a separation of $1.48$ astronomical units (au), close enough for the white dwarf to continually accrete material from its companion \cite{2016MNRAS.457..822B}.
At irregular intervals, enough material accumulates on the surface of the white dwarf to trigger a thermonuclear explosion, driving a quasi-spherical shock into the red giant's wind (see \citen{SOM}, Figure 2). Eight outbursts were observed between 1898 and 2006, recurring in intervals of $9\,$to$\,26$ years \cite{2009A&A...497..815B}.

\begin{figure}[ht!]%grid,tics
    \centering
    \includegraphics[width=0.49\textwidth]{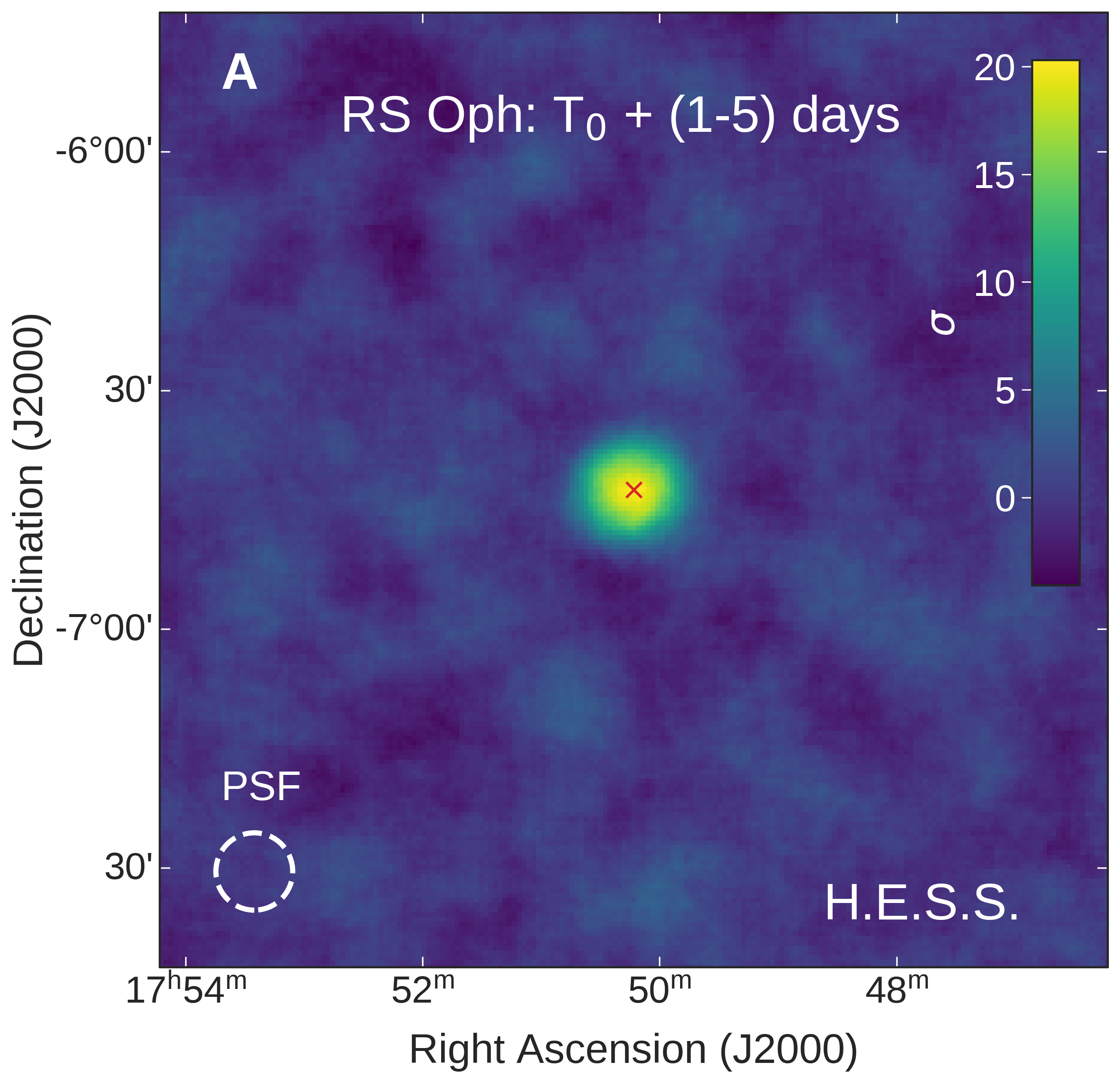}
    \includegraphics[width=0.49\textwidth]{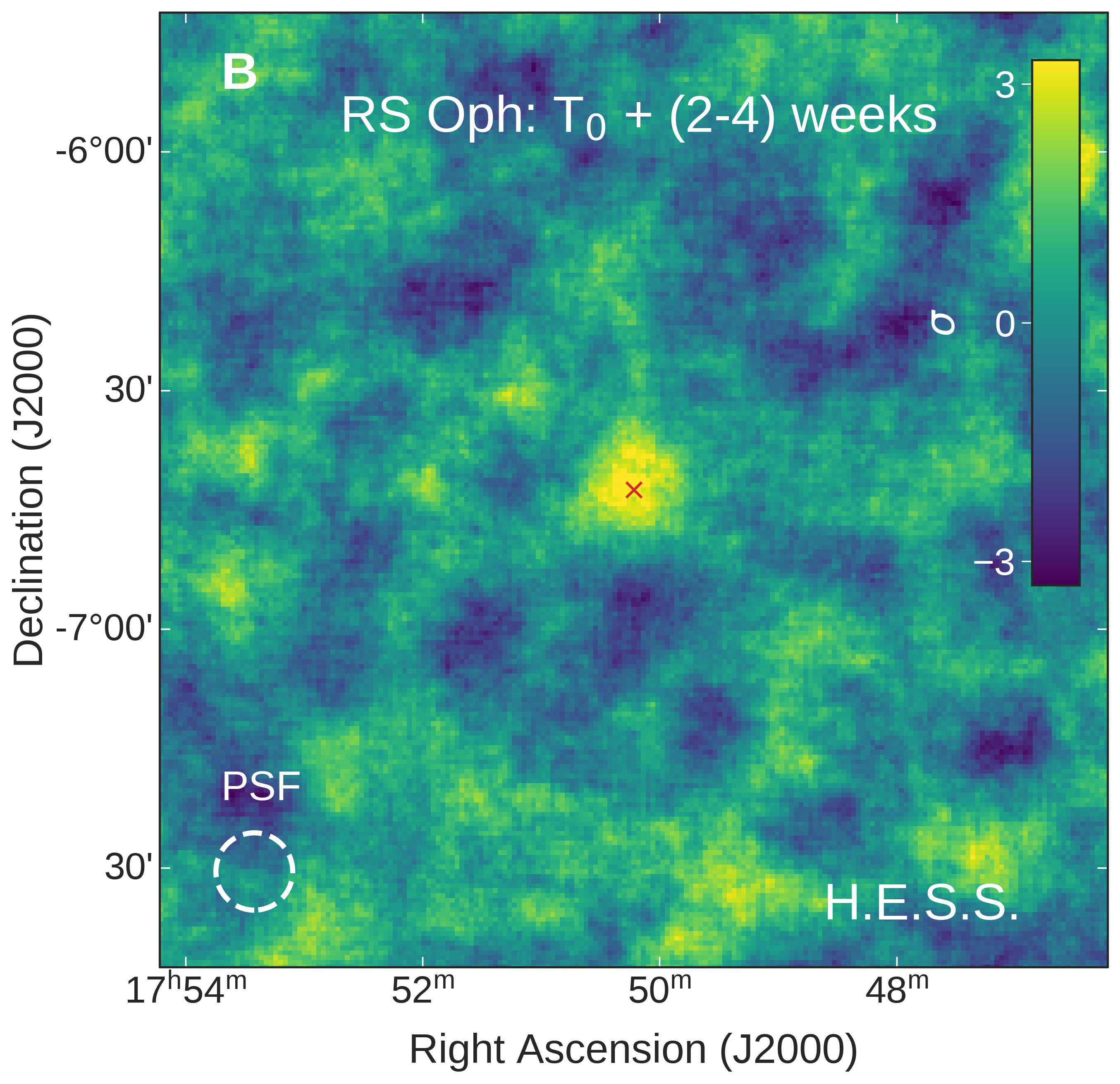}
    \caption{\textbf{RS Oph significance maps.} Significance maps derived from the H.E.S.S. $> 100$\,GeV gamma-ray observations for the early (A) and late (B) phases of the RS\,Oph 2021 outburst. T$_0=$ Modified Julian Day (MJD) 59435.25, is the time of peak optical emission. The dashed white circles indicate the point-spread-function (PSF).}
  \label{fig:skymap}
\end{figure}

On $8^{\rm th}$ August 2021, an outburst of RS\,Oph was identified from optical observations  \cite{aavso}, with a peak naked-eye visual magnitude of 4.5, over a thousand times brighter than the quiescent visual magnitude of 12.5. 
We observed RS Oph with the High Energy Stereoscopic System (H.E.S.S.), an array of five atmospheric Cherenkov telescopes. Observations commenced on $9^{\rm th}$ August 2021 and continued for five nights until $13^{\rm th}$ August 2021.
Optical background emission from the Moon prevented good quality observations for the following ten nights. 
During each of those five nights, H.E.S.S. detected point-like gamma-ray emission from the direction of RS\,Oph, with a significance of $>6$\,sigma on each night (see \citen{SOM}, Table S1). The  data for those five nights combined are shown in Figure~\ref{fig:skymap}. Observations recommenced on $25^{\rm th}$ August 2021, $\sim$17 days after the initial outburst. We find evidence for a much weaker signal ($\sim3 \sigma$ above the background) was seen in $\sim$15 hours (after quality cuts) of data accumulated over the following 14 days.

%integrated above X GeV

We performed a spectral analysis of the H.E.S.S. data for the first five observation nights separately. We also separated the data from the array of four $106\,\textrm{m}^2$ mirror area telescopes (designated CT1-4) and the fifth $612\,\textrm{m}^2$ mirror area  low-threshold telescope (CT5). We find that the VHE flux is variable, with a spectral index $>3$ throughout (see \citen{SOM}, Table S2). 

\begin{figure}[t!]
\centering
    \includegraphics[width=0.97\textwidth]{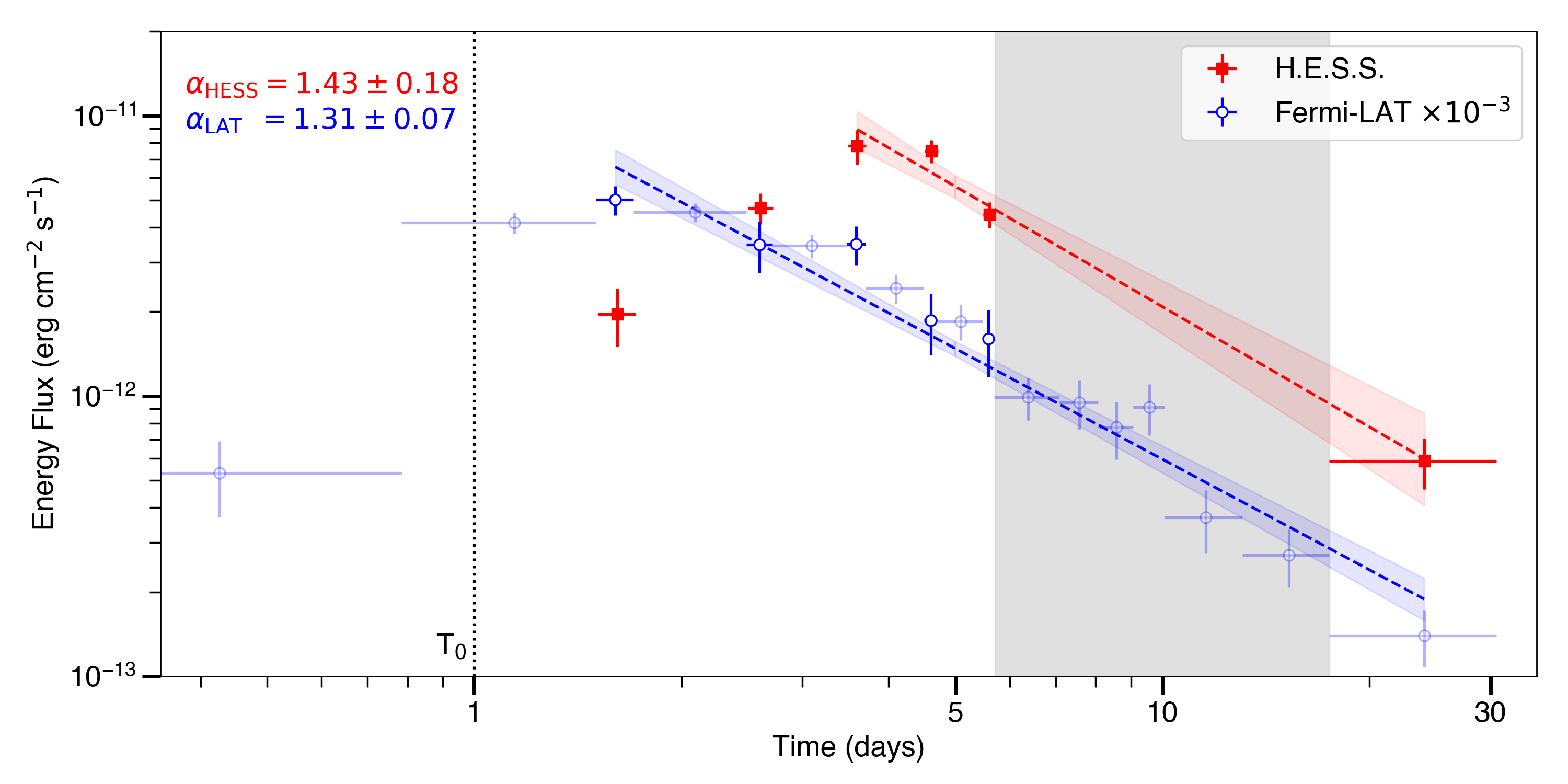}
      \caption{\textbf{Gamma-ray light curves of RS Oph.} Light curves of gamma-ray emission from RS\,Oph including data from \emph{Fermi}-LAT and H.E.S.S. observations. The H.E.S.S. data (red squares) cover a period of five nights, after which observations ceased for ten days due to bright moonlight, marked by the shaded grey band, then recommenced for a period of 14 days. The H.E.S.S. flux is integrated from  250\,GeV to 2.5\,TeV, whilst the \emph{Fermi}-LAT flux is integrated from  60\,MeV to 500\,GeV. \emph{Fermi}-LAT data are shown in 6-hour bins (blue circles) corresponding to the time windows of the H.E.S.S. observations, and data outside of these times shown with semi-transparent markers. Error bars are 1 $\sigma$ statistical uncertainties. A power-law slope model was fitted to the temporal decay after the time of peak flux for both instruments (red and blue dashed lines, with uncertainties indicated by the shaded regions). The vertical dotted black line indicates the peak of the outburst in the optical waveband, T$_0$.}
  \label{fig:lightcurve}
\end{figure}

Figure~\ref{fig:lightcurve} shows the time evolution of the gamma-ray flux curve for  photon energies between 250\,GeV and 2.5\,TeV. The VHE gamma-ray flux rises smoothly from T$_0$, the time of peak optical emission in the \textit{V} band \cite{aavso},
until a VHE peak on the third night of observations, after which the VHE gamma-ray energy flux decays by an order of magnitude over a two-week period. We obtained 60\,MeV -- 500\,GeV data taken by the \emph{Fermi}-LAT (Large Area Telescope) instrument for the same time period as the H.E.S.S. observations which are also shown in Figure~\ref{fig:lightcurve}. The flux varies in the range $\sim$1$\times$10$^{-8}$ -- 2$\times$10$^{-10}$ erg cm$^{-2}$s$^{-1}$, with a peak flux in the \emph{Fermi}-LAT data on $T_0 + 1$\,day. 
The VHE gamma-ray emission peak is delayed by a further two days. 

\begin{figure}[ht!]
    \includegraphics[width=0.5\textwidth]{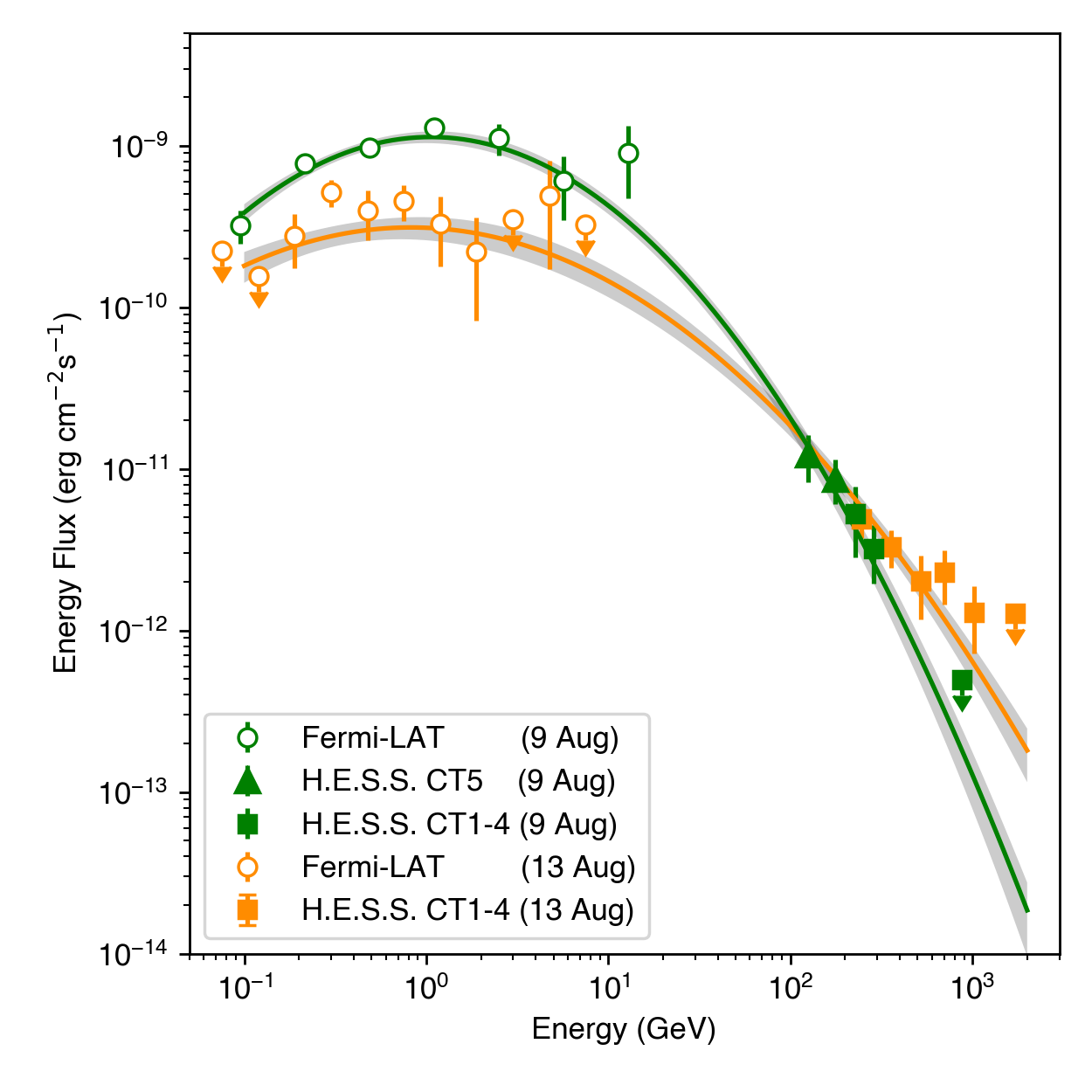}
    \caption{\textbf{RS Oph gamma-ray spectra.} The H.E.S.S. and \emph{Fermi}-LAT spectra for 9 (green) and 13 (orange) August fitted with a log-parabola model. 
    The analysis is applied separately for the H.E.S.S. CT1-4 (squares) and CT5 (triangles) - see text. The \emph{Fermi}-LAT 
    data (open circles) are integrated over $24$~h centred at the H.E.S.S. observation times. 
    There is clear spectral evolution from the 9$^{\rm th}$ to the 13$^{\rm th}$ August, with a noticeable reduction in the \emph{Fermi}-LAT flux as well as an increase in the maximum energy of the TeV spectrum.
    Error bars are 1 sigma statistical uncertainty, 
    and upper limits are the $95\%$ confidence level.
    }
  \label{fig:spectrum}
\end{figure}

After the peak flux, we fitted the decay in time $t$ of the energy flux with a power-law with exponent $\alpha$, $t^{-\alpha}$ and found best-fitting values of $\alpha \approx 1.3-1.4$ in both data sets: $\alpha_{\rm HESS}=1.43 \pm 0.18$ for H.E.S.S. and $\alpha_{\rm LAT}=1.31 \pm 0.07$ for \emph{Fermi}-LAT, for the choice of $T_0=$1\,day.
The \emph{Fermi}-LAT flux and temporal decay are consistent with that obtained from bins of 24-hour duration, the higher statistics of the larger bins enabling the detailed spectral analysis shown in Figure~\ref{fig:spectrum}.

The combined H.E.S.S. and \emph{Fermi}-LAT data \cite{SOM} allow us to measure wide-band gamma-ray spectra over more than four orders of magnitude in energy and follow their temporal evolution (Figure \ref{fig:spectrum}). The RS\,Oph spectra are consistent with a log-parabola model. Comparison of spectra taken on different nights show a general trend for the flux normalisation to decrease and the parabola to widen over time (see \citen{SOM}, Table S2).

%Discussion / Model
The similarity between the spectra of the \emph{Fermi}-LAT and H.E.S.S. data, and their similar decay profiles after their respective peaks, indicate a common origin for the gamma rays from one day to one month after the explosion. 
We assume that the particles that generate the gamma rays are accelerated at the external shock as it propagates into the wind of the red giant (\citen{SOM}, Figure S2). Optical spectroscopic measurements of the 2021 nova indicate shock velocities in the range  $u_{\rm sh} = 4000-5000$ \,km\,s$^{-1}$ \cite{2021ATel14852....1M}, compatible with measurements from the previous 2006 outburst of RS\,Oph \cite{2006ApJ...652..629B,2006Natur.442..276S}. 
High resolution images of the 2006 event \cite{2007ApJ...665L..63B} indicated the polar regions of the 
shock expanded at $\approx 5\,000$\,km\,s$^{-1}$ over the first 5 months.
We therefore assume that during the first week following the 2021 outburst the shock velocity did not fall below several thousand kilometers per second.

The images of the 2006 nova showed a quasi-spherical outflow, pinched at an equatorial ring \cite{2006Natur.442..279O,2007ApJ...665L..63B}. This is consistent with a shock expanding into the  wind of the red giant orthogonal to the orbital plane of the binary, but inhibited close to the plane by the denser gas  \cite{2016MNRAS.457..822B,2008A&A...484L...9W}. We expect particles to undergo diffusive shock acceleration at the external fast moving shocks, above and below the orbital plane of the binary. We consider two scenarios to explain the observed spectral and temporal properties of the gamma-ray emission from RS Oph: 
gamma-ray production from accelerated protons colliding with dense gas in the downstream volume (hadronic $\pi^0$ decay model), or gamma-ray emissions from energetic electrons scattering low energy photons from the nova (inverse Compton model).
For both models, the observations place strong constraints on the physical conditions, particularly the acceleration efficiencies required to match the measured fluxes and maximum photon energies \cite{SOM}.

VHE gamma-ray emission requires acceleration of particles to $>$TeV energies. 
The maximum energy a particle attains at a shock is determined by either the time taken before radiative cooling dominates over acceleration, or by when particles become too energetic and escape upstream of the shock \cite{Bell2013}. 
This confinement limit applies when the accelerating particles are unable to excite magnetic field fluctuations to a sufficient level ahead of the shock.
As particles spend longer diffusing upstream than downstream, the details of the downstream magnetic fields can be neglected \cite{1983A&A...125..249L}. For upstream magnetic-field amplification to be effective, a sufficient flux of particles, typically protons, must escape upstream of the shock. This requires both efficient transfer of the shock kinetic energy to relativistic protons, and that a fraction of these protons penetrate far upstream. Escaping particles have energies concentrated close to the maximum particle energy; less energetic particles are confined to the shock. 
The escaping flux per unit area at a given shock radius can be parameterized as 
\begin{equation}
    {j}_{\rm esc}= e \xi_{\rm esc} F_\varepsilon/ E_{\rm max}~,
\end{equation}
where $e$ is the elementary charge, $F_\varepsilon = \frac{1}{2}\rho_{\rm up} u_{\rm sh}^3$ is the energy flux density for a high Mach-number non-relativistic shock,  $\rho_{\rm up}$ is the immediate upstream gas density at that radius, $E_{\rm max}$ is the maximum particle energy which dominates the escaping flux, and the efficiency parameter $\xi_{\rm esc}$ depends on the assumed particle spectrum \cite{SOM}.
For a wind-like density profile, and neglecting radiative losses, the confinement limit on the maximum energy for a particle of charge $|Z|e$ (with atomic number $Z$) is 
\begin{equation}
    E_{\rm max} = 1.5 |Z| 
    \left(\frac{\xi_{\rm esc}}{0.01}\right)
    \left(\frac{\dot{M}/v_{\rm wind}}{10^{11}~{\rm kg~m}^{-1} }\right)^{1/2}
     \left(\frac{u_{\rm sh}}{5000~{\rm km~s}^{-1} }\right)^{2} ~{\rm TeV}\enspace,
\end{equation}
where $\dot{M}$ and $v_{\rm wind}$ are the mass-loss rate and the wind velocity of the red giant. $\xi_{\rm esc}$ is predicted to be about $1\%$ for high Mach-number shocks \cite{Bell2013}. 
For RS\,Oph,
$\dot{M}/v_{\rm wind}= 6\times10^{11}~{\rm kg~m}^{-1}$ \cite{2006Natur.442..279O} which, together with the inferred shock velocities, indicates a maximum energy $E_{\rm max} \approx 10$ TeV. This is compatible with the measured maximum photon energies $E_{\rm \gamma,max} \approx 1$\,TeV (Figure~\ref{fig:spectrum}). 

In the hadronic scenario, the gamma-ray lightcurves are consistent with an expanding shock in a decreasing density profile. With the adopted distance of 1.4\,kpc \cite{2008ASPC..401...52B}, the measured gamma-ray fluxes require that $>$10\% of the post-shocked medium's internal energy goes to accelerating protons or other nuclei. The delay between the peaks in the \emph{Fermi}-LAT and H.E.S.S. lightcurves would then reflect the finite acceleration time of the $>$1 TeV protons, or more specifically, the time taken to populate the high-energy tail of the distribution \cite{1991MNRAS.251..340D}. A simple calculation of the acceleration time, based on a comparison of the confinement and Hillas limits, implies that this should happen on the order of days \cite{SOM}.
This is consistent with the spectral evolution seen in Figure~\ref{fig:spectrum}: a reduction in the \emph{Fermi}-LAT flux, accompanied by a hardening in H.E.S.S. flux and increased $E_{\rm \gamma, max}$ over the first few days post-outburst. Attenuation of gamma-rays due to the novas optical and infra-red photon fields is  minor below 1\,TeV a few hours after the explosion, and therefore attenuation alone cannot account for the observed hardening (see \citen{SOM}, Figure S10).

For the alternative, leptonic scenario in which TeV gamma-rays are produced by VHE electrons, the acceleration needs to overcome the strong radiative losses due to inverse Compton cooling in the strong photon fields of the nova, as well as synchrotron cooling in the magnetic field in the shock region. To achieve this, electrons must accelerate at close to the Bohm rate, i.e. the scattering rate equal to the rate of gyration in the magnetic field \cite{1983A&A...125..249L}. Such efficient scattering requires strong self-generated magnetic fluctuations upstream of the shock, which implies the presence of an energetic relativistic hadronic component. 
In this scenario, the differences between the spectral slopes in the \emph{Fermi}-LAT and H.E.S.S. energy ranges are a consequence of the energy-dependent cooling rates in time-dependent photon fields. Electrons that radiate in the VHE band cool on a timescale less than the age of the nova remnant at the times the observations were taken, while lower-energy un-cooled electrons accumulate downstream over time. The \emph{Fermi}-LAT light curve in this scenario then reflects the evolution of the energy density of soft-photon targets, while the H.E.S.S. lightcurve traces the full radiative output of high-energy electrons up to the VHE peak. After the peak, due to the rapidly decreasing photon energy density, the cooling time increases faster than the remnant's age, and the VHE emitting electrons are also slow cooling.

From the time-dependent numerical single-zone model, parameters can be found to approximately describe the light curves and spectra in both leptonic and hadronic scenarios, providing quantitative estimates for the acceleration efficiencies at early times $t$ ($t\leq T_0+5$ days). We are unable to account for the temporal decay at later times since a more complex treatment of the internal structure of the nova remnant, its non-spherical geometry, and escape of particles from the emission zone is required.
Both the leptonic and hadronic models at early times are consistent with continuous injection of particles following a power law spectrum in energy $E^{-2.2}$  with a high-energy cut-off. To approximately match the observed flux, the hadronic model requires $>$10\% of the shocked gas' internal energy to be transferred to non-thermal protons, while the leptonic model requires $>$1\% efficiency for non-thermal electrons. 

Such a high fraction of the total energy in non-thermal electrons is inconsistent with theories of injection at high Mach-number shocks, for which ion injection efficiency is expected to be much higher than that of electrons \cite{MalkovDrury}. Numerical simulations of high Mach-number shocks find a ratio of $\ll$10$^{-2}$ for electron to ion energy densities \cite{2015PhRvL.114h5003P}, which is consistent with multi-wavelength models of supernova remnants \cite{2008A&A...483..529V}. A $>$1\% efficiency of conversion to non-thermal electrons cannot be realised in a purely leptonic model.
For this reason, we prefer the hadronic scenario discussed above, for which both the implied high proton acceleration efficiencies and inferred maximum energy are in line with theoretical predictions \cite{Bell2013}. Our findings support previous hadronic models of gamma-ray novae \cite{2016MNRAS.457.1786M,2017NatAs...1..697L,2020NatAs...4..776A}.\\

The VHE detection of RS Oph demonstrates that particle acceleration to $\sim$TeV energies can occur within the dense wind environments of recurrent novae.  The total kinetic energy from each nova of RS Oph is estimated to be $\sim 10^{43}$\,erg
($10^{-7}$\,solar masses of ejecta at $\sim$4000\,km\,s$^{-1}$ \cite{2006Natur.442..276S}),
with a large fraction of this being converted to relativistic protons and heavier nuclei, which are the main constituents of Galactic cosmic rays. Each nova event generates enough cosmic rays to fill a cubic parsec (pc) volume with an energy density of $\sim$0.1\,eV\,cm$^{-3}$, similar to the local Galactic cosmic-ray energy density of $\sim(0.8-1.0)$\,eV\,cm$^{-3}$ \cite{Cummings2016} sustained by supernovae.
In the case of RS\,Oph, the cosmic-ray energy input recurs approximately every $\Delta t = 15-20$ years,
leading to an almost continuous injection of non-thermal particles. 
For a diffusion coefficient $D$ in the neighbourhood of RS Oph, the cosmic-ray output from each nova is spread over a diffusion length $\ell_{\rm diff}=\sqrt{4D~\Delta t}$. Using a Galactic average $D\approx (3-5) \times 10^{28}$ cm$^2$ s$^{-1}$ \cite{2007ARNPS..57..285S}, we find $\ell_{\rm diff}> 1$\,pc, and the contribution from each nova is sub-dominant to the average Galactic cosmic-ray population.
If the diffusion coefficient in the neighbourhood of the nova is much lower than the Galactic average, due to enhanced turbulence following previous novae for example, such a sustained source of cosmic rays will raise the local abundance. If efficient acceleration of particles to TeV energies in recurrent novae is commonplace, with spectral energy distribution harder than that of the Galactic cosmic-ray background $\propto E^{-2.7}$, the local contribution from novae would dominate at TeV energies over volumes $ > {\rm pc}^3$. The size of the affected region will depend on the value of the diffusion coefficient, which can be constrained using measurements of the diffuse gamma-ray emission at energies $\sim10-100$ GeV \cite{FermiLAT2012}.

Our time-resolved gamma-ray emission measurements have implications for the origin of cosmic rays. Acceleration of cosmic rays to PeV energies in supernova remnants requires substantial amplification of magnetic fields. Fast shocks (\(\sim10,000\)\,km\,s$^{-1}$) propagating through the dense winds ($\dot{M}/v_{\rm wind}\sim 10^{13}~{\rm kg~m}^{-1}$) associated with the progenitors of supernova remnants from massive ($\gtrsim 8\,\mathrm{M}_\odot$) stars, provide the only known environments where the required conditions can (in theory) be met \cite{Bell2013, Marcowith2018}. However, observational confirmation of this prediction has not been found. 
The detection of VHE gamma-rays from RS\,Oph provides an example of a Galactic accelerator reaching the theoretical limit for the maximum achievable particle energy via diffusive shock acceleration \cite{Bell2013}. If our results can be extrapolated to the most optimistic supernova conditions, they support the prevailing model of Galactic PeV cosmic rays originating in supernova remnants from massive stars \cite{Bell2013, Marcowith2018}.

\newpage

\nocite{Crab2006}
%TC:ignore
\bibliographystyle{science}
\bibliography{main.bib}

\begin{thebibliography}{10}

\bibitem{2010Sci...329..817A}
A.~A. {Abdo}, {\it et~al.\/}, {\it Science\/} {\bf 329}, 817 (2010).

\bibitem{2014Sci...345..554A}
M.~{Ackermann}, {\it et~al.\/}, {\it Science\/} {\bf 345}, 554 (2014).

\bibitem{ChomiukReview}
L.~Chomiuk, B.~D. Metzger, K.~J. Shen, {\it Annual Review of Astronomy and
  Astrophysics\/} {\bf 59}, 391 (2021).

\bibitem{2008ASPC..401...52B}
R.~K. {Barry}, {\it et~al.\/}, {\it RS Ophiuchi (2006) and the Recurrent Nova
  Phenomenon\/}, A.~{Evans}, M.~F. {Bode}, T.~J. {O'Brien}, M.~J. {Darnley},
  eds. (2008), vol. 401 of {\it Astronomical Society of the Pacific Conference
  Series\/}, p.~52.

\bibitem{2018A&A...616A...1G}
{Gaia Collaboration}, {\it et~al.\/}, {\it \aap\/} {\bf 616}, A1 (2018).

\bibitem{2021A&A...649A...4L}
L.~{Lindegren}, {\it et~al.\/}, {\it \aap\/} {\bf 649}, A4 (2021).

\bibitem{2016MNRAS.457..822B}
R.~A. {Booth}, S.~{Mohamed}, P.~{Podsiadlowski}, {\it \mnras\/} {\bf 457}, 822
  (2016).

\bibitem{SOM}
{Materials and methods are available as supplementary materials.}

\bibitem{2009A&A...497..815B}
E.~{Brandi}, C.~{Quiroga}, J.~{Miko{\l}ajewska}, O.~E. {Ferrer}, L.~G.
  {Garc{\'\i}a}, {\it \aap\/} {\bf 497}, 815 (2009).

\bibitem{aavso}
S.~Kafka, {Observations from the AAVSO International Database},
  \url{https://www.aavso.org} (2021).

\bibitem{2021ATel14852....1M}
J.~{Mikolajewska}, E.~{Aydi}, D.~{Buckley}, C.~{Galan}, M.~{Orio}, {\it The
  Astronomer's Telegram\/} {\bf 14852}, 1 (2021).

\bibitem{2006ApJ...652..629B}
M.~F. {Bode}, {\it et~al.\/}, {\it \apj\/} {\bf 652}, 629 (2006).

\bibitem{2006Natur.442..276S}
J.~L. {Sokoloski}, G.~J.~M. {Luna}, K.~{Mukai}, S.~J. {Kenyon}, {\it \nat\/}
  {\bf 442}, 276 (2006).

\bibitem{2007ApJ...665L..63B}
M.~F. {Bode}, {\it et~al.\/}, {\it \apjl\/} {\bf 665}, L63 (2007).

\bibitem{2006Natur.442..279O}
T.~J. {O'Brien}, {\it et~al.\/}, {\it \nat\/} {\bf 442}, 279 (2006).

\bibitem{2008A&A...484L...9W}
R.~{Walder}, D.~{Folini}, S.~N. {Shore}, {\it \aap\/} {\bf 484}, L9 (2008).

\bibitem{Bell2013}
A.~R. {Bell}, K.~M. {Schure}, B.~{Reville}, G.~{Giacinti}, {\it \mnras\/} {\bf
  431}, 415 (2013).

\bibitem{1983A&A...125..249L}
P.~O. {Lagage}, C.~J. {Cesarsky}, {\it \aap\/} {\bf 125}, 249 (1983).

\bibitem{1991MNRAS.251..340D}
L.~O. {Drury}, {\it \mnras\/} {\bf 251}, 340 (1991).

\bibitem{MalkovDrury}
M.~A. {Malkov}, L.~O. {Drury}, {\it Reports on Progress in Physics\/} {\bf 64},
  429 (2001).

\bibitem{2015PhRvL.114h5003P}
J.~{Park}, D.~{Caprioli}, A.~{Spitkovsky}, {\it \prl\/} {\bf 114}, 085003
  (2015).

\bibitem{2008A&A...483..529V}
H.~J. {V{\"o}lk}, E.~G. {Berezhko}, L.~T. {Ksenofontov}, {\it \aap\/} {\bf
  483}, 529 (2008).

\bibitem{2016MNRAS.457.1786M}
B.~D. {Metzger}, {\it et~al.\/}, {\it \mnras\/} {\bf 457}, 1786 (2016).

\bibitem{2017NatAs...1..697L}
K.-L. {Li}, {\it et~al.\/}, {\it Nature Astronomy\/} {\bf 1}, 697 (2017).

\bibitem{2020NatAs...4..776A}
E.~{Aydi}, {\it et~al.\/}, {\it Nature Astronomy\/} {\bf 4}, 776 (2020).

\bibitem{Cummings2016}
A.~C. {Cummings}, {\it et~al.\/}, {\it \apj\/} {\bf 831}, 18 (2016).

\bibitem{2007ARNPS..57..285S}
A.~W. {Strong}, I.~V. {Moskalenko}, V.~S. {Ptuskin}, {\it Annual Review of
  Nuclear and Particle Science\/} {\bf 57}, 285 (2007).

\bibitem{FermiLAT2012}
M.~{Ackermann}, {\it et~al.\/}, {\it \apj\/} {\bf 750}, 3 (2012).

\bibitem{Marcowith2018}
A.~{Marcowith}, V.~V. {Dwarkadas}, M.~{Renaud}, V.~{Tatischeff}, G.~{Giacinti},
  {\it \mnras\/} {\bf 479}, 4470 (2018).

\bibitem{Crab2006}
F.~{Aharonian}, {\it et~al.\/}, {\it Astronomy and Astrophysics\/} {\bf 457},
  899 (2006).

\bibitem{2015arXiv150902902H}
M.~{Holler}, {\it et~al.\/}, {\it PoS\/} {\bf ICRC2015}, 847 (2016).

\bibitem{Ashton2020}
T.~{Ashton}, {\it et~al.\/}, {\it Astroparticle Physics\/} {\bf 118}, 102425
  (2020).

\bibitem{FC2021Technical}
B.~{Bi}, {\it et~al.\/}, {\it Proceedings of 37th International Cosmic Ray
  Conference — PoS(ICRC2021)\/}  (2021).

\bibitem{2021ATel14834....1C}
C.~C. {Cheung}, S.~{Ciprini}, T.~J. {Johnson}, {\it The Astronomer's
  Telegram\/} {\bf 14834}, 1 (2021).

\bibitem{2021ATel14838....1T}
K.~{Taguchi}, T.~{Ueta}, K.~{Isogai}, {\it The Astronomer's Telegram\/} {\bf
  14838}, 1 (2021).

\bibitem{2021ATel14840....1M}
U.~{Munari}, P.~{Valisa}, {\it The Astronomer's Telegram\/} {\bf 14840}, 1
  (2021).

\bibitem{Hahn2014}
J.~{Hahn}, {\it et~al.\/}, {\it Astroparticle Physics\/} {\bf 54}, 25 (2014).

\bibitem{Berge2007}
D.~{Berge}, S.~{Funk}, J.~{Hinton}, {\it Astronomy and Astrophysics\/} {\bf
  466}, 1219 (2007).

\bibitem{Ohm2009}
S.~Ohm, C.~van Eldik, K.~Egberts, {\it Astroparticle Physics\/} {\bf 31}, 383
  (2009).

\bibitem{Murach2015}
T.~{Murach}, M.~{Gajdus}, R.~D. {Parsons}, {\it {Proceedings of the 34th
  International Cosmic Ray Conference (ICRC2015), The Hague, The
  Netherlands}\/} (2015).

\bibitem{FC2021}
G.~{P{\"u}hlhofer}, {\it et~al.\/}, {\it Proceedings of 37th International
  Cosmic Ray Conference — PoS(ICRC2021)\/}  (2021).

\bibitem{Parsons2014}
R.~D. {Parsons}, J.~A. {Hinton}, {\it Astroparticle Physics\/} {\bf 56}, 26
  (2014).

\bibitem{Mohrmann2019}
L.~{Mohrmann}, {\it et~al.\/}, {\it \aap\/} {\bf 632}, A72 (2019).

\bibitem{Nigro2019}
C.~{Nigro}, {\it et~al.\/}, {\it \aap\/} {\bf 625}, A10 (2019).

\bibitem{Deil2017}
C.~{Deil}, {\it et~al.\/}, {\it 35th International Cosmic Ray Conference
  (ICRC2017)\/} (2017), vol. 301 of {\it International Cosmic Ray
  Conference\/}, p. 766.

\bibitem{deNaurois2009}
M.~{de Naurois}, L.~{Rolland}, {\it Astroparticle Physics\/} {\bf 32}, 231
  (2009).

\bibitem{RXJ2006}
F.~{Aharonian}, {\it et~al.\/}, {\it \aap\/} {\bf 449}, 223 (2006).

\bibitem{Atwood2013}
W.~B. {Atwood}, {\it et~al.\/}, {\it \apj\/} {\bf 774}, 76 (2013).

\bibitem{Wood2017}
M.~{Wood}, {\it et~al.\/}, {\it 35th International Cosmic Ray Conference
  (ICRC2017)\/} (2017), vol. 301 of {\it International Cosmic Ray
  Conference\/}, p. 824.

\bibitem{2019ascl.soft05011F}
{Fermi Science Support Development Team}, {Fermitools: Fermi Science Tools}
  (2019).

\bibitem{4fgl_2020ApJS..247...33A}
S.~{Abdollahi}, {\it et~al.\/}, {\it \apjs\/} {\bf 247}, 33 (2020).

\bibitem{FSSC}
{\url{http://fermi.gsfc.nasa.gov/ssc/data/access/lat/BackgroundModels.html}}.

\bibitem{2020ApJ...905...62A}
E.~{Aydi}, {\it et~al.\/}, {\it \apj\/} {\bf 905}, 62 (2020).

\bibitem{2006ApJ...653L.141D}
R.~{Das}, D.~P.~K. {Banerjee}, N.~M. {Ashok}, {\it \apjl\/} {\bf 653}, L141
  (2006).

\bibitem{1985MNRAS.217..205B}
M.~F. {Bode}, F.~D. {Kahn}, {\it \mnras\/} {\bf 217}, 205 (1985).

\bibitem{1959book...Sedov}
L.~I. {Sedov}, {\it {Similarity and Dimensional Methods in Mechanics}\/}
  (Academic Press, New York, 1959).

\bibitem{2017MNRAS.465.3793T}
X.~{Tang}, R.~A. {Chevalier}, {\it \mnras\/} {\bf 465}, 3793 (2017).

\bibitem{LandauFluids}
L.~D. {Landau}, E.~M. {Lifshitz}, {\it {Fluid mechanics}\/} (Pergamon Press
  Ltd., 1959).

\bibitem{1992MNRAS.255..683O}
T.~J. {O'Brien}, M.~F. {Bode}, F.~D. {Kahn}, {\it \mnras\/} {\bf 255}, 683
  (1992).

\bibitem{1998ApJ...501..339P}
D.~{Proga}, S.~J. {Kenyon}, J.~C. {Raymond}, {\it \apj\/} {\bf 501}, 339
  (1998).

\bibitem{1999isw..book.....L}
H.~J.~G.~L.~M. {Lamers}, J.~P. {Cassinelli}, {\it {Introduction to Stellar
  Winds}\/} (Cambridge University Press, 1999).

\bibitem{2021arXiv210901101M}
U.~{Munari}, P.~{Valisa}, {\it arXiv e-prints\/} p. arXiv:2109.01101 (2021).

\bibitem{Drury1983}
L.~O. {Drury}, {\it Reports on Progress in Physics\/} {\bf 46}, 973 (1983).

\bibitem{2008MNRAS.386..509R}
B.~{Reville}, S.~{O'Sullivan}, P.~{Duffy}, J.~G. {Kirk}, {\it \mnras\/} {\bf
  386}, 509 (2008).

\bibitem{Parker1958}
E.~N. {Parker}, {\it \apj\/} {\bf 128}, 664 (1958).

\bibitem{Kemball1997}
A.~J. {Kemball}, P.~J. {Diamond}, {\it \apjl\/} {\bf 481}, L111 (1997).

\bibitem{Hillas}
A.~M. {Hillas}, {\it \araa\/} {\bf 22}, 425 (1984).

\bibitem{Bell2004}
A.~R. {Bell}, {\it \mnras\/} {\bf 353}, 550 (2004).

\bibitem{BlumenthalGould}
G.~R. {Blumenthal}, R.~J. {Gould}, {\it Reviews of Modern Physics\/} {\bf 42},
  237 (1970).

\bibitem{2007ApJ...671L.157E}
A.~{Evans}, {\it et~al.\/}, {\it \apjl\/} {\bf 671}, L157 (2007).

\bibitem{Kafexhiu2014}
E.~{Kafexhiu}, F.~{Aharonian}, A.~M. {Taylor}, G.~S. {Vila}, {\it \prd\/} {\bf
  90}, 123014 (2014).

\end{thebibliography}
%\bibliography{som}

\section*{Acknowledgements}

\noindent {\bf Acknowledgements} 
% The collaboration acknowledgements are available on the website linked to this publication: \url{https://www.mpi-hd.mpg.de/hfm/HESS/pages/publications/}. 
%https://www.mpi-hd.mpg.de/hfm/HESS/pages/publications/auxiliary/HESS-Acknowledgements-2021.html
We appreciate the excellent work of the technical
support staff in Berlin, Zeuthen, Heidelberg, Palaiseau, Paris,
Saclay, Tübingen and in Namibia in the construction and operation
of the equipment. This work benefited from services provided by the
H.E.S.S. Virtual Organisation, supported by the national resource
providers of the EGI Federation.

\noindent {\bf Funding}
The support of the Namibian authorities and of the University of
Namibia in facilitating the construction and operation of H.E.S.S.
is gratefully acknowledged, as is the support by the German
Ministry for Education and Research (BMBF), the Max Planck Society,
the German Research Foundation (DFG), the Helmholtz Association,
the Alexander von Humboldt Foundation, the French Ministry of
Higher Education, Research and Innovation, the Centre National de
la Recherche Scientifique (CNRS/IN2P3 and CNRS/INSU), the
Commissariat à l’énergie atomique et aux énergies alternatives
(CEA), the U.K. Science and Technology Facilities Council (STFC),
the Irish Research Council (IRC) and the Science Foundation Ireland
(SFI), the Knut and Alice Wallenberg Foundation, the Polish
Ministry of Education and Science, agreement no. 2021/WK/06, the
South African Department of Science and Technology and National
Research Foundation, the University of Namibia, the National
Commission on Research, Science \& Technology of Namibia (NCRST),
the Austrian Federal Ministry of Education, Science and Research
and the Austrian Science Fund (FWF), the Australian Research
Council (ARC), the Japan Society for the Promotion of Science, the
University of Amsterdam and the Science Committee of Armenia grant
21AG-1C085.

\noindent {\bf Author Contributions:}
A.~Mitchell and S.~Ohm led the H.E.S.S.  observations of RS\,Oph. R. Konno carried out initial on-site analysis, the main H.E.S.S. CT1-4 stereo data analysis and the atmospheric correction. S. Steinmassl performed the CT5 mono analysis, and J.P. Ernenwein the cross-check analysis. S. Ohm coordinated the multiple H.E.S.S. analyses and evaluated the systematic errors. E. de O\~na Wilhelmi and T. Unbehaun carried out the \emph{Fermi}-LAT data analysis. B. Reville, D. Khangulyan, J.Mackey and E. de O\~na Wilhelmi developed the interpretation and modelling. The manuscript was prepared by A. Mitchell, B. Reville, S. Ohm, D. Khangulyan, E. de O\~na Wilhelmi, R. Konno, and S. Steinmassl. S. Wagner is the collaboration spokesperson. Other H.E.S.S. collaboration authors contributed to the design, construction and operation of H.E.S.S., the development and maintenance of data handling, data reduction or data analysis software. All authors meet the journal’s authorship criteria and have reviewed, discussed, and commented on the results and the manuscript.

\noindent {\bf Competing Interests:} The authors declare that they have no competing interests.

\noindent {\bf Data and Materials Availability:} The H.E.S.S. data are available at: \url{https://www.mpi-hd.mpg.de/hfm/HESS/pages/publications/auxiliary/2022_RS_Oph/}. This includes the sky maps (cf. Figure~1), the light-curve (cf. Figure~2), the points of the spectral energy distributions (cf. Figure~3). 

\textbf{Supplementary Material: Authors and affiliations, Materials and Methods, Figure~S1-S10, Tables S1-S4, References (30-71)}%62

\newpage

\baselineskip24pt

\setcounter{page}{0}
\begin{center}
\includegraphics[width=0.6\textwidth]{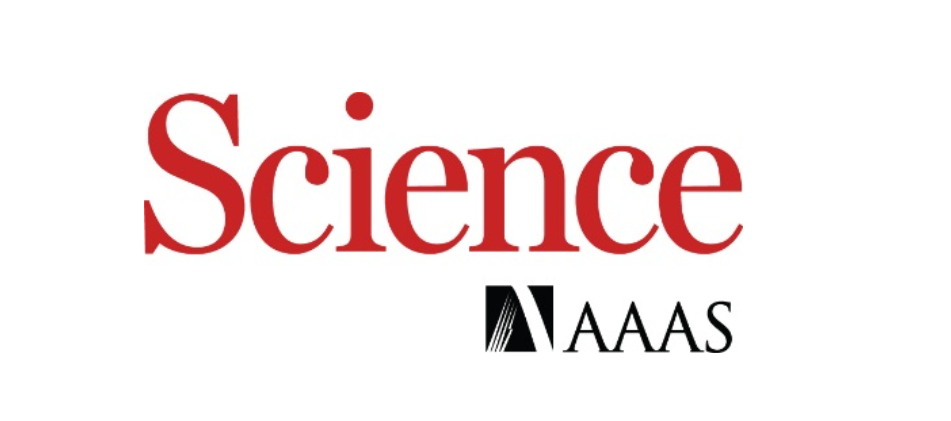}
\end{center}

\bigskip
\begin{center}
{\LARGE Supplementary Material for}\\
\bigskip
{\Large Time-resolved hadronic particle acceleration in the recurrent nova RS\,Ophiuchi}\\
\bigskip
{\Large H.E.S.S. Collaboration*}\\
\medskip
{\par *Correspondence to: contact.hess@hess-experiment.eu; Alison Mitchell (alison.mw.mitchell@fau.de), Stefan Ohm (stefan.ohm@desy.de), Brian Reville (brian.reville@mpi-hd.mpg.de)}
\end{center}

\noindent
{\large This PDF file includes:}

\bigskip

\begin{spacing}{1.0}
\noindent
\hspace*{1.2cm} Materials and Methods\\
\hspace*{1.2cm} Supplementary Text \\
\hspace*{1.2cm} Figs. S1 to S10\\
\hspace*{1.2cm} Tables S1 to S4
\end{spacing}

\newpage
\subsection*{H.E.S.S. Collaboration authors and affiliations}
%\author[]{
F.~Aharonian$^{1,2,3}$,
F.~Ait~Benkhali$^{4}$,
E.O.~Ang\"uner$^{5}$,
H.~Ashkar$^{6}$,
M.~Backes$^{7,8}$,
V.~Baghmanyan$^{9}$,
V.~Barbosa~Martins$^{10}$,
R.~Batzofin$^{11}$,
Y.~Becherini$^{12,13}$,
D.~Berge$^{10}$
% ...
\\
1. Dublin Institute for Advanced Studies, 31 Fitzwilliam Place, Dublin 2, Ireland \\
2. Max-Planck-Institut f\"ur Kernphysik, P.O. Box 103980, D 69029 Heidelberg, Germany \\
3. High Energy Astrophysics Laboratory, Russian-Armenian University (RAU),  123 Hovsep Emin St  Yerevan 0051, Armenia \\
4. Landessternwarte, Universit\"at Heidelberg, K\"onigstuhl, D 69117 Heidelberg, Germany \\
5. Aix Marseille Universit\'e, Centre national de la recherche scientifique (CNRS)/Institut National de Physique Nucl\'eaire et Physique des Particules (IN2P3), Centre de Physique des Particules de Marseille (CPPM), Marseille, France \\
6. Laboratoire Leprince-Ringuet, École Polytechnique, CNRS, Institut Polytechnique de Paris, F-91128 Palaiseau, France \\
7. University of Namibia, Department of Physics, Private Bag 13301, Windhoek 10005, Namibia \\
8. Centre for Space Research, North-West University, Potchefstroom 2520, South Africa \\
9. Instytut Fizyki J\c{a}drowej Polskiej Akademii Nauk (PAN), ul. Radzikowskiego 152, 31-342 Krak{\'o}w, Poland \\
10. Deutsches Elektronen-Synchrotron DESY, Platanenallee 6, 15738, Germany \\
11. School of Physics, University of the Witwatersrand, 1 Jan Smuts Avenue, Braamfontein, Johannesburg, 2050 South Africa \\
12. Université de Paris, CNRS, Astroparticule et Cosmologie, F-75013 Paris, France \\
13. Department of Physics and Electrical Engineering, Linnaeus University,  351 95 V\"axj\"o, Sweden \\
14. Institut f\"ur Astronomie und Astrophysik, Universit\"at T\"ubingen, Sand 1, D 72076 T\"ubingen, Germany \\
15. Laboratoire Univers et Théories, Observatoire de Paris, Université PSL, CNRS, Université de Paris, 92190 Meudon, France \\
16. Sorbonne Universit\'e, Universit\'e Paris Diderot, Sorbonne Paris Cit\'e, CNRS/IN2P3, Laboratoire de Physique Nucl\'eaire et de Hautes Energies (LPNHE), 4 Place Jussieu, F-75252 Paris, France \\
17. Université Savoie Mont Blanc, CNRS, Laboratoire d'Annecy de Physique des Particules - IN2P3, 74000 Annecy, France \\
18. Institute for Research on the Fundamental Laws of the Universe (IRFU), Commisariat à l'énergie atomique (CEA), Universit\'e Paris-Saclay, F-91191 Gif-sur-Yvette, France \\
19. University of Oxford, Department of Physics, Denys Wilkinson Building, Keble Road, Oxford OX1 3RH, UK \\
20. Institut f\"ur Physik und Astronomie, Universit\"at Potsdam,  Karl-Liebknecht-Strasse 24/25, D 14476 Potsdam, Germany \\
21. School of Physical Sciences, University of Adelaide, Adelaide 5005, Australia \\
22. Friedrich-Alexander-Universit\"at Erlangen-N\"urnberg, Erlangen Centre for Astroparticle Physics, Erwin-Rommel-Str. 1, D 91058 Erlangen, Germany \\
23. Laboratoire Univers et Particules de Montpellier, Universit\'e Montpellier, CNRS/IN2P3,  CC 72, Place Eug\`ene Bataillon, F-34095 Montpellier Cedex 5, France \\
24. Universit\'e Bordeaux, CNRS, Laboratoire de Physique des Deux Infinis (LP2i), Bordeaux, Joint Research Unit (UMR 5797), F-33170 Gradignan, France \\
25. Institut f\"ur Astro- und Teilchenphysik, Leopold-Franzens-Universit\"at Innsbruck, A-6020 Innsbruck, Austria \\
26. Universit\"at Hamburg, Institut f\"ur Experimentalphysik, Luruper Chaussee 149, D 22761 Hamburg, Germany \\
27. Obserwatorium Astronomiczne, Uniwersytet Jagiello{\'n}ski, ul. Orla 171, 30-244 Krak{\'o}w, Poland \\
28. Institute of Astronomy, Faculty of Physics, Astronomy and Informatics, Nicolaus Copernicus University,  Grudziadzka 5, 87-100 Torun, Poland \\
29. Department of Physics, Rikkyo University, 3-34-1 Nishi-Ikebukuro, Toshima-ku, Tokyo 171-8501, Japan \\
30. Nicolaus Copernicus Astronomical Center, Polish Academy of Sciences, ul. Bartycka 18, 00-716 Warsaw, Poland \\
31. Institut f\"ur Physik, Humboldt-Universit\"at zu Berlin, Newtonstr. 15, D 12489 Berlin, Germany \\
32. Department of Physics and Astronomy, The University of Leicester, University Road, Leicester, LE1 7RH, United Kingdom \\
33. Gravitation and Astroparticle Physics at the University of Amsterdam (GRAPPA), Anton Pannekoek Institute for Astronomy, University of Amsterdam,  Science Park 904, 1098 XH Amsterdam, The Netherlands \\
34. Yerevan Physics Institute, 2 Alikhanian Brothers St., 375036 Yerevan, Armenia \\
35. Kavli Institute for the Physics and Mathematics of the Universe (World Premier International Research Center Initiative (WPI)), The University of Tokyo Institutes for Advanced Study (UTIAS), The University of Tokyo, 5-1-5 Kashiwa-no-Ha, Kashiwa, Chiba, 277-8583, Japan \\
36. Department of Physics, Konan University, 8-9-1 Okamoto, Higashinada, Kobe, Hyogo 658-8501, Japan \\
37. Institute of Physical and Chemical Research (RIKEN), 2-1 Hirosawa, Wako, Saitama 351-0198, Japan \\
38. Department of Physics, University of the Free State,  PO Box 339, Bloemfontein 9300, South Africa \\
%}

\renewcommand{\thetable}{S\arabic{table}}
\renewcommand{\thefigure}{S\arabic{figure}}
\renewcommand{\theequation}{S\arabic{equation}}

\setcounter{figure}{0}
% \newcounter{mybibstartvalue}
% \setcounter{mybibstartvalue}{30}

\newpage

\section*{Materials and methods}

\subsection*{H.E.S.S. Observations and Data Analysis}

H.E.S.S. is an array of five Imaging Atmospheric Cherenkov Telescopes located in the Khomas Highland of Namibia, sensitive to VHE gamma rays in the energy range between a few tens of GeV and a few tens of TeV \cite{Crab2006,2015arXiv150902902H}. H.E.S.S. consists of four 12\,m-diameter Davies-Cotton telescopes (CT1-4), equipped with Cherenkov cameras \cite{Ashton2020} and have a field-of-view (FoV) of 5$^\circ$. The largest telescope (CT5) is located in the middle of the CT1-4 array, has a 28\,m-diameter mirror and is equipped with a Cherenkov camera with 3.4$^\circ$ FoV \cite{FC2021Technical}. Observations of classical novae by H.E.S.S. are triggered by external alerts and follow-up observations are decided on a case-by-case basis taking into account if a transient signal is reported by \emph{Fermi}-LAT, if the ejecta velocity is high ($\gtrsim 1500\,\mathrm{km\,s}^{-1}$), and if the peak optical magnitude is bright (m$_\mathrm{V}$\,$\lesssim 9$). H.E.S.S. follows up on $\sim$two nova candidates on average per year, subject to them satisfying at least one of the aforementioned conditions. 
Initial reports of an outburst of RS\,Oph on $8^{\rm th}-9^{\rm th}$ August 2021 satisfied all three conditions, with a 6\,$\sigma$ detection by \emph{Fermi}-LAT \cite{2021ATel14834....1C}, ejecta velocity of $\gtrsim 2600\,\mathrm{km\,s}^{-1}$ \cite{2021ATel14838....1T, 2021ATel14840....1M} and optical magnitude of m$_\mathrm{V}$\,$\sim 5.0$, prompting H.E.S.S. observations with high priority. 
RS\,Oph was the first nova visible to H.E.S.S. to meet all three conditions at the same time. 

Follow-up observations of RS\,Oph with H.E.S.S. started on August 9$^{\rm{th}}$, 18:17:40 (Coordinated Universal Time, UTC) and concluded on September 7$^{\rm{th}}$, 19:47:31 (UTC). H.E.S.S. accumulated a total of 24.3 hours of observations over the first five nights after the explosion. After the full moon period, during which H.E.S.S. observations are paused, H.E.S.S. continued RS\,Oph observations for a further 32.9 hours. The observations were conducted in the standard ``wobble mode'' where telescopes are pointed at alternate offsets of 0.5$^\circ$ from the position of RS\,Oph (right ascension 17$^{\mathrm{h}}$50$^{\mathrm{min}}$13.16$^{\mathrm{s}}$, declination -06$^{\circ}$42$'$28.5$''$ (J2000 equinox)). To achieve a low energy threshold, we only selected data for analysis with zenith angle $\lesssim 35^\circ$, with a resulting average zenith angle of $22^\circ$.

Throughout the first five nights, the data set is subject to strongly variable observing conditions. Due to increasing levels of moonlight, a spectral analysis of CT5 data is performed for the first three nights only. On the second and third nights, atmospheric conditions were poor due to a higher-than-usual aerosol content in the atmosphere, whilst atmospheric conditions were good on the first, fourth and fifth night. We applied corrections to account for the night-to-night variations in atmospheric conditions (see below). Furthermore, we conducted observations under low to moderate moonlight during nights 2 to 5, which increased the noise from night-sky-background (NSB) photons by a factor of 1.5 in night 4 and 2.5 in night 5 compared to dark sky conditions (c.f. Table~\ref{tab:hess_obs}). 

\begin{table}[t!]
    \centering
    \begin{tabular}{ccccccc}
    Night & $T_{\mathrm{obs}}$ & Livetime & Significance & Atmospheric transparency & NSB noise level & Telescopes \\
     & (UTC) & (hours) & ($\sigma$) & & & \\\hline
    09 Aug. 2021 & 18:17:40 & 3.2 & 5.8 (6.4) & 0.90 & 1.0 & CT1-5\\
    10 Aug. 2021 & 17:53:46 & 3.7 (2.8) & 9.0 (7.1) & 0.80 & 1.0 & CT1-5\\
    11 Aug. 2021 & 17:44:08 & 3.7 & 9.8 (9.6) & 0.65 & 1.0 & CT1-5\\
    12 Aug. 2021 & 18:17:12 & 2.3 & 13.6 & 1.00 & 1.5 & CT1-4\\
    13 Aug. 2021 & 17:44:43 & 2.8 & 10.5 (9.4) & 1.10 & 2.5 & CT1-5\\\hline
    25 Aug. -- 07 Sep. 2021 & 17:48:03; 19:47:31 & 14.6 (13.4) & 3.3 (2.3) & 0.96 & 1.0 & CT1-5\\
    \end{tabular}
    \caption{{\bf Summary of H.E.S.S. observations and data set used in the analysis.} The start time $T_{\mathrm{obs}}$ and duration of observations is given for each night. Values are for the CT1-4 stereo analysis, except numbers in brackets which denote CT5 mono analysis results. The atmospheric transparency is quantified using the Cherenkov transparency coefficient \cite{Hahn2014}. The NSB noise level is given relative to the nominal Galactic level experienced by the telescopes. Only observations taken under good atmospheric conditions and low zenith angle $\lesssim 35^\circ$ were used in the data set from 25 Aug. 2021 to 07 Sep. 2021, leaving $\sim$15 hours for analysis out of a total of 32.9 hours of observations.}
    \label{tab:hess_obs}
\end{table}

We performed the analysis of the H.E.S.S. data using both events recorded by CT1-4 in stereoscopic (stereo) mode, and in a monoscopic (mono) analysis of CT5. The ring-background method \cite{Berge2007} was employed for the extraction of skymaps for both analyses. The resulting significance maps are shown in Figure 1 for CT1-4. Background events were classified and rejected using multivariate analysis techniques \cite{Ohm2009, Murach2015, FC2021}. The reconstruction of shower properties was performed using a pixel-based maximum likelihood technique (CT1-4 stereo, \cite{Parsons2014}) and a multivariate-based reconstruction (CT5 mono, \cite{Murach2015, FC2021}). The spectral reconstruction and model fitting were performed with a likelihood-based procedure, using the \textsc{Gammapy} software package \cite{Mohrmann2019, Nigro2019, Deil2017} version 0.18.2. In the case of the CT5 mono analysis, the reflected-region \cite{Berge2007} method was used to define background control regions and to derive spectral results.

Significant VHE gamma-ray emission was detected by H.E.S.S. from the direction of RS\,Oph on each of the five nights from $9^{\rm th}-13^{\rm th}$ August 2021 (Table \ref{tab:hess_obs}) in an online analysis and later confirmed offline by two independent analysis chains \cite{Ohm2009,Parsons2014,deNaurois2009}. The cross-check analysis employs an independent calibration, reconstruction and background suppression \cite{deNaurois2009}. 
On night 4, CT5 did not participate in observations for technical reasons, but it detected RS\,Oph in VHE gamma rays in the other four nights. 

RS\,Oph was observable again with H.E.S.S. under good low mooonlight conditions on August $25^{\rm{th}}$ and observations continued as long as the source was visible to H.E.S.S. until September 7$^{\mathrm{th}}$.
The analysis of these observations revealed a VHE gamma-ray signal from RS\,Oph at the $3 - 4\sigma$-level in the main CT1-4 analysis as well as in the cross-check analysis. No significant emission was detected with the CT5 mono analysis in this late phase, consistent with its somewhat lower sensitivity. The corresponding flux upper limit is $F_\mathrm{UL}(E>250\,\mathrm{GeV}) = 3.7 \times10^{-12}\,\mathrm{cm}^{-2}\,\mathrm{s}^{-1}$.
Table~\ref{tab:hess_obs} summarizes the H.E.S.S. observations considered in our analysis below.

\begin{figure}[t]
    \centering
    \includegraphics[width = 0.8\linewidth]{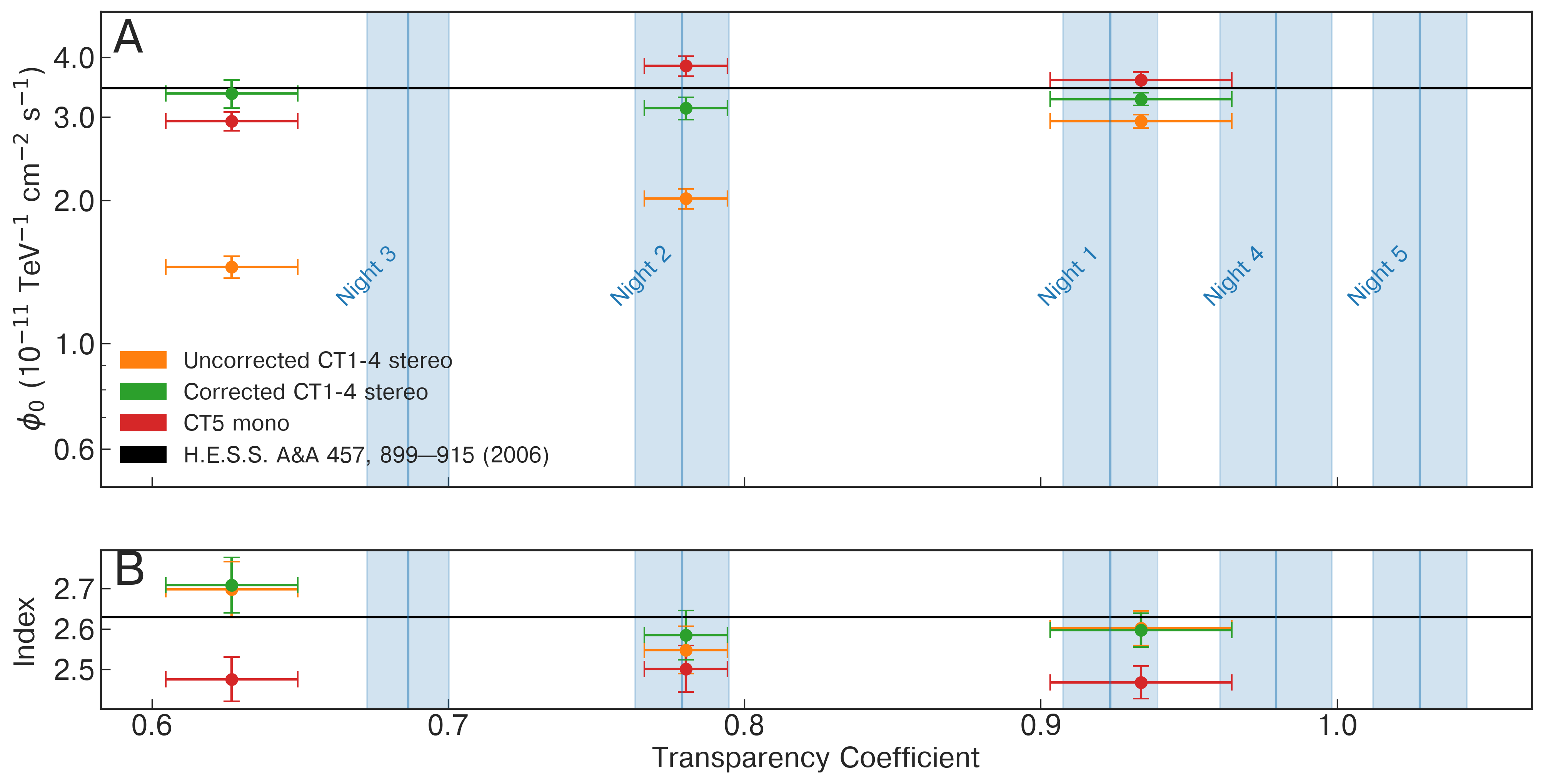}
    \caption{\textbf{Atmospheric transparency correction of Crab Nebula observations.} Illustration of the effect of atmospheric transparency correction to Crab Nebula observations taken during the same observing season as RS\,Oph for a range of atmospheric transparency conditions. The top panel illustrates the impact of the correction on the normalization $\phi_0$ of the reconstructed flux, $\phi = \phi_0 \cdot (E/1\,{\rm TeV})^{-\Gamma}$. The bottom panel shows the reconstructed index $\Gamma$ without and with correction. Vertical blue lines indicate the atmospheric transparency conditions corresponding to the five nights of RS\,Oph observations.}
    \label{fig:crab}
\end{figure}

The quality of the atmospheric conditions is quantified using the Cherenkov transparency coefficient \cite{Hahn2014} with lower values corresponding to lower transmission of Cherenkov light through the atmosphere, as reported in Table \ref{tab:hess_obs} for the RS\,Oph observations. To correct for the different atmospheric conditions and resulting light-yield of the telescopes, the energy scale of the energy migration matrix as well as the effective areas have been scaled for individual observations with the atmospheric transparency coefficient for the CT1-4 stereo analysis. To validate this correction, the reconstructed Crab Nebula spectrum (after scaling the energy scale of the instrument response functions (IRFs) with the transparency coefficient) is compared to the published H.E.S.S. spectrum, using the same IRFs as employed for the RS\,Oph observations. A good agreement between the spectral parameters of the corrected Crab Nebula spectrum and a previously published spectrum \cite{Crab2006} is found for a range of atmospheric conditions, with the flux matching to within 10\% (see Figure~\ref{fig:crab}.)

For the CT5 mono analysis, two different sets of IRFs tailored to the different atmospheric conditions have been used. A comparison study to the same Crab Nebula data set as described above was performed for the CT5 mono analysis, confirming that also here the Crab Nebula flux and gamma-ray spectral index is reconstructed correctly for the range of atmospheric transparencies experienced during the RS\,Oph observations (Figure~\ref{fig:crab}).

To test for spectral variation during the outburst, the H.E.S.S. data were analysed in nightly bins for the five nights with significant detection (see Table \ref{tab:hess_obs}) and fitted with a power-law model. Best-fitting parameters are given in Table \ref{tab:night_spec} for the first three nights using the CT5 mono analysis and for all five nights using the CT1-4 stereo analysis. Nightly spectral results are consistent between the two configurations.

\begin{table}[t!]
    \centering
    \begin{tabular}{c c|c c c}
     &Data set & $\phi_0$ & $E_0$  & Index $\Gamma$ \\
     &&[$10^{-11}$ TeV$^{-1}$ cm$^{-2}$ s$^{-1}$] & [TeV]\\
     \hline
     mono &09 Aug. 2021 &$14.9 \pm (2.7)_\text{stat.} \pm (3.0)_\text{syst.}$  &  0.18 & $ 3.22  \pm (0.38)_\text{stat.} \pm (0.20)_\text{syst.}$ \\
     &10 Aug. 2021 &$25.2 \pm (4.7)_\text{stat.} \pm (5.0)_\text{syst.}$  & 0.18 & $ 4.01 \pm (0.48)_\text{stat.} \pm (0.20)_\text{syst.}$ \\
     &11 Aug. 2021 &$28.5 \pm (3.3)_\text{stat.} \pm (5.7)_\text{syst.}$  & 0.18 & $ 3.15 \pm (0.23)_\text{stat.} \pm (0.20)_\text{syst.}$ \\     
     stereo &09 Aug. 2021 & $0.91 \pm (0.28)_\text{stat.} \pm (0.14)_\text{syst.}$  & 0.35 & $4.24 \pm (0.75)_\text{stat.} \pm (0.15)_\text{syst.}$ \\
     &10 Aug. 2021 & $1.90 \pm (0.32)_\text{stat.} \pm (0.38)_\text{syst.}$  &  0.35 & $3.32 \pm (0.30)_\text{stat.} \pm (0.15)_\text{syst.}$ \\
     &11 Aug. 2021 & $3.57 \pm (0.54)_\text{stat.} \pm (0.54)_\text{syst.}$  &  0.35 & $4.08 \pm (0.42)_\text{stat.} \pm (0.20)_\text{syst.}$ \\
     &12 Aug. 2021 & $3.00 \pm (0.33)_\text{stat.} \pm (0.45)_\text{syst.}$   &  0.35 & $3.27 \pm (0.21)_\text{stat.} \pm (0.15)_\text{syst.}$ \\
     &13 Aug. 2021 & $1.77 \pm (0.25)_\text{stat.} \pm (0.35)_\text{syst.}$  &  0.35 & $3.24 \pm (0.24)_\text{stat.} \pm (0.15)_\text{syst.}$ \\ 
          \hline
     stereo & 25 Aug. 2021 - 07 Sep. 2021 & $0.238\pm (0.080)_\text{stat.} \pm (0.036)_\text{syst.}$  &  0.35 & $3.33 \pm (0.45)_\text{stat.} \pm (0.15)_\text{syst.}$ 
    \end{tabular}
    \caption{ \textbf{Nightly spectral measurements of RS\,Oph.} Best-fitting nightly spectral parameters from H.E.S.S. assuming a power-law model of the form $\phi_0\left(\frac{E}{E_0}\right)^{-\Gamma}$, with derived systematic uncertainties. $E_0$ is the reference energy, $\phi_0$ is the amplitude at the reference energy and $\Gamma$ is the spectral index.}
    \label{tab:night_spec}
\end{table}

\subsubsection*{Systematic Uncertainties}
\label{section:systematics}
Multiple sources of systematic uncertainties contribute to affect the spectral measurements of the CT1-4 stereo and the CT5 mono analyses. We summarise the various sources and their estimated influence on the reconstructed flux and gamma-ray spectral index. Systematic uncertainties stemming from Monte-Carlo extensive air shower hadronic interaction models, broken pixels of the Cherenkov cameras, and the live time of the data set are sub-dominant for both analyses, but have been taken into account \cite{Crab2006}.

For the CT1-4 analysis, two different sets of cuts to select gamma-ray-like events have been applied to the night 1-5 data set, showing a systematic difference in the reconstructed flux normalisation and spectral index of $\sim$5\% and 0.1, respectively. The tests conducted using Crab Nebula observations under varying atmospheric conditions capture run-by-run differences in atmospheric conditions after correction, differences in the assumed and actual optical telescopes efficiency, and the scaling to correct for low atmospheric transparency itself. This results in a systematic error of the flux normalisation of 10\%. The impact on the reconstructed spectral index varies from 0.05 (good, medium) to 0.15 (poor) atmospheric transparency. An additional 10\% systematic uncertainty of the flux normalisation is assumed for nights 2 and 5, when the Cherenkov cameras were operated at higher camera trigger settings and during moonlight.

The uncertainty in the CT5 mono analysis is derived from the following aspects. Following \cite{RXJ2006}, an imperfect description of the background acceptance leads to a systematic uncertainty of $\sim$15\% on the flux normalisation and 0.15 on the spectral index for the CT5 mono analysis. The Crab Nebula spectral comparison under comparable atmospheric conditions yields a systematic error on the flux normalisation of 5\%, 10\% and 15\% for good, medium and poor atmospheric conditions, respectively. The uncertainty on the spectral index is at a level of 0.15 for the different transparencies. 

A comparison to the intrinsic spectrum of the Crab Nebula \cite{Nigro2019}, shows that the overall energy scale uncertainty of the H.E.S.S. observations is at the 15\% level. Table~\ref{tab:night_spec} includes the resulting systematic errors of the RS\,Oph spectral measurements for nights 1 to 5 and the CT1-4 stereo as well as the CT5 mono analysis. The systematic error for the late follow-up observations is derived in the same way to that described above.

\subsection*{\emph{Fermi}-LAT Data Analysis}
We analyse \emph{Fermi}-LAT data to investigate spectral evolution in bins of 24 hours, centred on 20:00\,UTC over the duration of the H.E.S.S. observations. We use \emph{Fermi}-LAT (P8R3, pass 8 release 3 \cite{Atwood2013}) data spanning from 9$^{\rm th}$ August 2021 to 7$^{\rm th}$ September 2021, 
in the energy range from 60\,MeV to 500\,GeV. We retrieve the data from a region of interest (ROI) defined by a radius of $15^\circ$ around the position of RS\,Oph. The analysis of the LAT data described above is performed using the $\textsc{fermipy}$ $\textsc{python}$ package (version 0.18.0), based on the $\textsc{Fermi Science Tools}$ (version 2.0.8)
\cite{Wood2017,2019ascl.soft05011F}. We analyzed SOURCE class events with a maximum zenith angle of
$90^\circ$
to eliminate Earth limb events. To derive the daily spectrum, the data are binned in 8 energy bins per decade and spatial bins of $0.1^\circ$ size. (The re-binned points in Figure 3 in the main article were determined only for plotting purposes, after the best-fitting parameters had been found.) The response of the LAT instrument is evaluated with the IRFs, version P8R3\_SOURCE\_V2) and we include in the model of the region all the LAT sources listed in the \emph{Fermi}-LAT Fourth Source Catalog (4FGL, \cite{4fgl_2020ApJS..247...33A}) in a radius of $20^\circ$ around the position of the nova. The contributions from Galactic and extra-galactic diffuse gamma-rays are described using the Galactic (gll\_iem\_v07) and isotropic (iso\_P8R3\_SOURCE\_V2\_v1) diffuse emission models. The models are available from the Fermi Science Support Center (FSSC) \cite{FSSC}. The energy dispersion correction is applied to all individual models describing the diffuse and discrete gamma-ray emission in the ROI, except for the isotropic diffuse emission model.

Spectral results obtained using 6 hour bins contemporaneous to the H.E.S.S. observations are found to be compatible with the results of the 24 hour bins, albeit with higher uncertainty. Additionally, we evaluate the systematic errors for the daily observations by applying the same analysis to a pre-flare 24~hour observation window. No significant gamma-ray signal is found at the position of the RS\,Oph nova in the pre-flare time interval, with a signal first detected during the optical rise time. This demonstrates that neither systematic effects in the Galactic diffuse emission model nor source confusion impact the \emph{Fermi}-LAT flux measurement. 

To compute the light curve, the time range is split into a series of bins of varying duration. For the first five nights of the H.E.S.S. observations, contemporaneous 6\,h bins are used (and 18\,h bins in between these times). Outside of these five nights, we use bins of 24\,h duration (or multiples thereof) as appropriate, given the statistics. 
To calculate the energy flux for a given time window, we model the sources' gamma-ray emission with a log-parabola spectral function of the form $\phi(E) = \phi_0 \left( \frac{E}{E_0}\right)^{-\alpha-\beta\log\left(E/E_0\right)}$, with parameters set to a spectral index $\alpha=2.0$, curvature$ \beta=0.13$ and reference energy $E_0=1$\,GeV, respectively. The flux normalisation $\phi_0$ is left free to vary to obtain the integrated flux in the 60\,MeV to 500\,GeV energy range.

We use the derived \emph{Fermi}-LAT spectral points for each observation period to perform joint spectral fits with the H.E.S.S. mono and stereo data, using the \textsc{Gammapy} software package (version 0.18.2) \cite{Deil2017}. The different data sets were fitted simultaneously using a log-parabola spectral model; the best fitting parameters for each night are quoted in Table \ref{tab:joint_spec}.
The general trend is for the flux normalisation to decrease and the parabola to widen, over the course of the five nights.

\begin{table}[ht!]
    \centering
    \begin{tabular}{c|cccc}
       Data set  & $\phi_0$ & $E_0$  & $\alpha$ & $\beta$ \\
      & [$10^{-4}$ TeV$^{-1}$ cm$^{-2}$ s$^{-1}$] & [GeV] & & \\
       \hline
    09 Aug. 2021 & $7.06 \pm 0.58$ & 1.0 & $1.98 \pm 0.04$ & $0.19 \pm 0.01$ \\ 
    10 Aug. 2021 & $4.27 \pm 0.43$ & 1.0 & $2.12 \pm 0.05$ & $0.13 \pm 0.01$ \\ 
    11 Aug. 2021 & $3.69 \pm 0.42$ & 1.0 & $2.01 \pm 0.06$ & $0.13 \pm 0.01$ \\ 
    12 Aug. 2021 & $1.79 \pm 0.32$ & 1.0 & $2.02 \pm 0.09$ & $0.11 \pm 0.02$ \\ 
    13 Aug. 2021 & $1.94 \pm 0.32$ & 1.0 & $2.05 \pm 0.07$ & $0.12 \pm 0.02$ 
    \end{tabular}
    \caption{\textbf{Nightly log-parabola fits for RS Oph.} Best-fitting spectral parameters from a joint fitting of the \emph{Fermi}-LAT and H.E.S.S. data with a log-parabola model. $E_0$ is the reference energy, $\phi_0$ is the amplitude at the reference energy, $\alpha$ is the spectral index and $\beta$ is the curvature.}%these values have been updated
    \label{tab:joint_spec}
\end{table}

\section*{Supplementary Text}

\subsection*{Modelling of gamma-ray emission}
\label{modelling}

Recurrent novae are complex hydrodynamic processes accompanied by the formation of multiple shock waves. The shocks are either internal, i.e., formed by collision of different parts of ejecta, or external, i.e., formed in the circumbinary medium. Both internal and external shocks might be responsible for the production of gamma-ray emission. While internal shocks remain a hypothesis (see, however, \cite{2020ApJ...905...62A}), external shock formation is unavoidable and the emission produced at the external shock is expected to be long-lasting. This was evident from the multi-wavelength observations of the 2006 RS Oph nova event \cite{2006ApJ...652..629B, 2006Natur.442..276S, 2006Natur.442..279O,2006ApJ...653L.141D,2007ApJ...665L..63B}. We therefore focus on the external shock component. Similarly to supernova explosions, the external shocks of recurrent novae have different phases of evolution \cite{1985MNRAS.217..205B}. Initially, when the circumbinary medium cannot substantially influence the explosion dynamics, the blast wave
expands with approximately constant velocity,  the so-called  ejecta-dominated phase. Once a sufficient amount of material is swept up, the shock decelerates, transitioning to the Sedov-Taylor phase. The shock deceleration rate further increases in the radiative phase. For a point-like explosion in a radial wind with a mass density ($\rho$) profile $\rho(r)\propto r^{-2}$,
the shock radius ($R_{\rm sh}$) satisfies $R_{\rm sh} \propto t^{2/3}$ during the Sedov-Taylor phase\cite{1985MNRAS.217..205B}. When cooling dominates the shock enters the radiative phase, with $R_{\rm sh} \propto t^{1/2}$.
Observations of the previous 2006 outburst of RS Oph indicate that after day $\approx6$ the spectroscopic data is consistent with a shock speed $u_{sh}\propto t^{-0.6}$ \cite{2006ApJ...652..629B,2006ApJ...653L.141D}.
While $u_{sh}\propto t^{-0.3}$ over the first 3 weeks, it is inconsistent with the expected X-ray flux \cite{2006Natur.442..276S}, indicating a self-similar solution does not apply. 
Follow-up imaging of the expanding shock from the 2006 outburst \cite{2007ApJ...665L..63B} showed that the polar expansion continued at about $5\,500$\,km\,s$^{-1}$ for at least 155 days, apparently with minimal deceleration.
This is consistent with 3D hydrodynamical simulations showing late-time expansion at $>\,4\,000$\,km\,s$^{-1}$ in the polar directions \cite{2016MNRAS.457..822B}.
This points to the potential complexity of the source geometry and difficulty in interpreting the measurements in the context of a model with a single shock velocity.
The mid-latitude regions with slower shocks, denser gas and stronger radiative cooling, could dominate the spectral measurements.
Assuming a similar evolutionary profile for the 2021 outburst, the source was detected with H.E.S.S. over the first five days primarily during the ejecta-dominated phase, during which the velocity is in the range $(2-4)\times 10^3$\,km\,s$^{-1}$. The velocity may have decreased by the time H.E.S.S. observations recommenced after the bright moonlight break.
Optical spectroscopy observations
suggest that on day 2 the shock was faster than found in earlier measurements made on day 1 \cite{2021ATel14840....1M,2021ATel14852....1M}. This we attribute to the shock propagation in an inhomogeneous stellar wind, as found in previous 3D simulations \cite{2016MNRAS.457..822B}.

\begin{figure}[t]
    \centering
    \includegraphics[width = 0.75\linewidth]{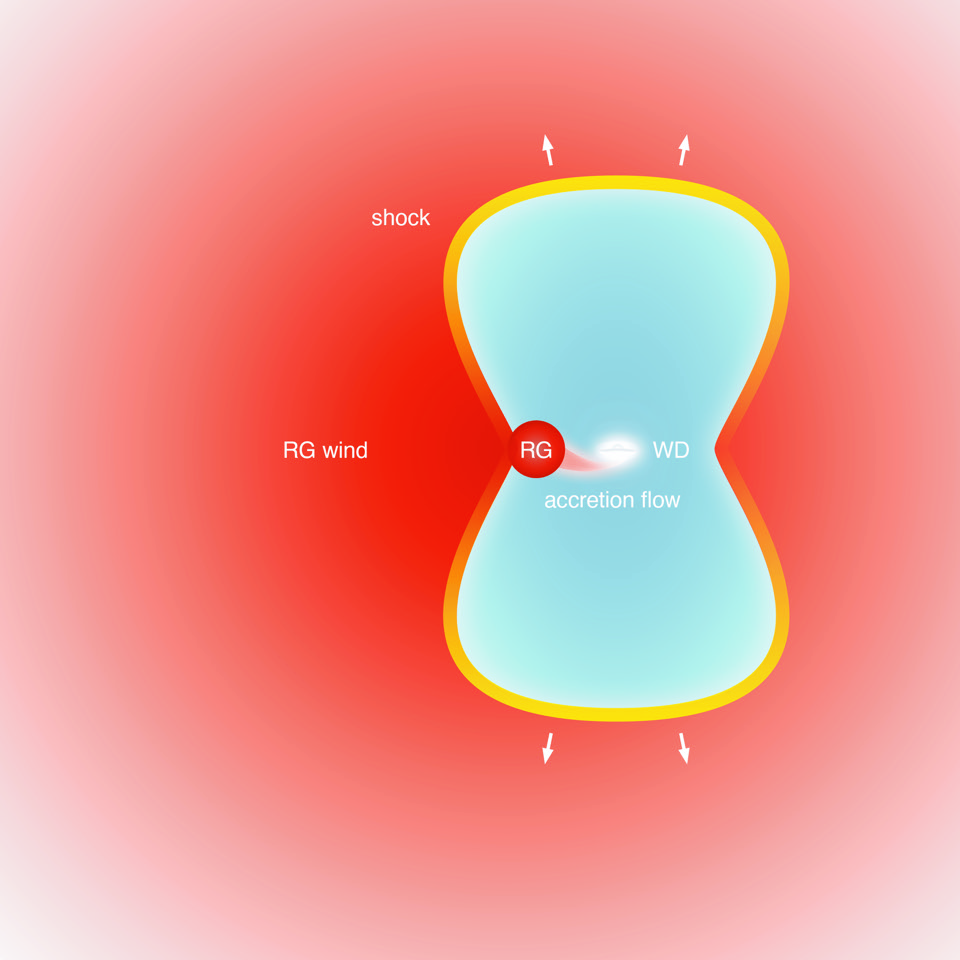}
    \caption{\textbf{Schematic of external shock model.} The explosion originates near the surface of the white dwarf (WD). Within one day, the shock is expanding as a bipolar blast wave moving orthogonal to the accretion disk, into the wind of the red giant (RG). The colour gradient of the shock indicates the expansion velocity, whilst the red gradient of the surrounding medium indicates the density of the RG wind. Material internal to the shock is shown in blue. Figure is not to scale.}
    \label{fig:cartoon}
\end{figure}

\paragraph{Shock conditions:}
We assume that the explosion origin is located at the white dwarf star, which moves around a red giant star on an orbit of
radius \(r_{\rm orb}\approx 1.48\rm \,au\) \cite{2016MNRAS.457..822B}. The accretion onto the white dwarf proceeds through a dense disk in the
orbital plane of the system. This results in the quasi-spherical blast wave initially moving faster in directions perpendicular to
the orbital plane, as two sections of a spherical shock, as illustrated in Figure \ref{fig:cartoon}. This picture is consistent with observations of the previous
nova event of RS Oph \cite{2006Natur.442..279O}. 
The period relevant for the H.E.S.S. observations likely covers the transition of the outflow expansion from the bipolar to quasi-spherical regimes (and also from the ejecta-dominated to the Sedov-Taylor phases), thus self-similar solutions (see, e.g. \cite{1959book...Sedov,2017MNRAS.465.3793T}) cannot describe the entirety of this period consistently. 

We employ a simple numerical model that describes the motion of the ejecta (namely distance covered by the ejecta \(R_{\mathrm{ej}}\), and its speed \(v_{\mathrm{ej}}\)) and the dynamics of the explosion's forward shock (its radius \(R_{\mathrm{sh}}\), and speed \(u_{\mathrm{sh}}\)). These quantities are obtained under the following assumptions: (i) the density and pressure across the shocked region are constant, and determined by the Rankine–Hugoniot conditions \cite{LandauFluids};  (ii) the upstream medium density is
determined by the distance covered by the shock as
\begin{equation}
\rho_{\mathrm{up}} = \frac{\dot{M}}{4\pi v_{\rm w} \left(r_{\rm orb}^2+R_{\mathrm{sh}}^2\right)}\,.
\label{eq:S1}
\end{equation}
Here \(\dot{M}\) is the mass-loss rate of the red giant star and \(v_{\rm w}\) is its wind speed, which depends on the distance as
\begin{equation}\label{eq:beta_speed}
    v_{\rm w}=v_\infty\left(1 -\frac{R_*}{\sqrt{r_{\rm orb}^2+R_{\mathrm{sh}}^2}}\right)^{\beta_{\rm w}}\,,
\end{equation}
with $v_\infty$ the terminal wind speed.
We take the value \(R_*\approx0.35\rm\,au\) for the radius of the red giant \cite{1992MNRAS.255..683O} and adopt for the exponent of the wind acceleration \(\beta_{\rm w}=3\) \cite{1998ApJ...501..339P}. For the red giant star in
RS\,Oph the wind parameter \(\dot{M}/v_{\infty}\) is estimated to be in the range \((5.7~-~7.4)\times10^{11}~{\rm kg~m}^{-1}\) \cite{1992MNRAS.255..683O}. We therefore adopt a fiducial value of \(\dot{M}/v_{\infty}= 6\times10^{11}~{\rm kg~m}^{-1}\).
For a typical terminal wind velocity of \(v_\infty=(10-30)\rm\,km\,s^{-1}\) for red giants \cite{1999isw..book.....L}, 
 this corresponds to a mass-loss rate of $\dot{M}\approx (1-3)\times10^{-7}M_\odot\rm\,yr^{-1}$ 
(where \(M_\odot\) is the solar mass). 

\begin{figure}[ht]
    \centering
    \includegraphics[width=\textwidth]{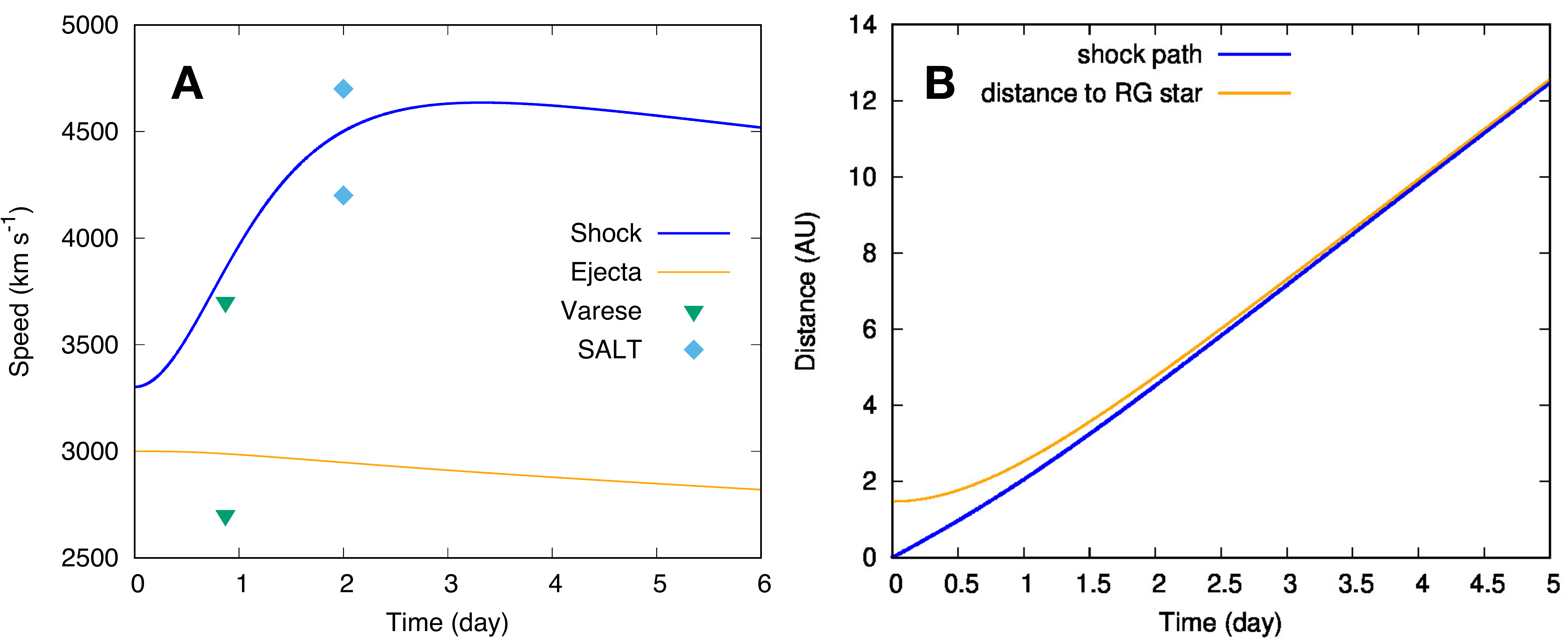}
    \caption{\textbf{Temporal shock evolution.} \textbf{A:} Model shock speed as a function of time since the explosion. Markers correspond to the expansion velocity estimated from the absorption center and the blue absorption edge of the \(H\alpha\) and \(H\beta\) P Cyg profiles obtained with Varese \cite{2021ATel14840....1M,2021arXiv210901101M} and SALT \cite{2021ATel14852....1M} spectroscopy of RS Oph. The shock speed is obtained in the framework of a single-zone model for initial ejecta velocity \(v_{\mathrm{ej},0}=3\,000\rm\,km\,s^{-1}\). \textbf{B:} distance covered by the forward shock and distance between the center of the RG star and forward shock. }
    \label{fig:shock_speed}
\end{figure}

For a strong shock in single-atom polytropic gas, one can estimate the shocked gas ($\rho_{\mathrm{ds}}$) mass-density as
\begin{equation}\label{eq:rho_down}
\rho_{\mathrm{ds}} \approx 10^{-14}\left(\frac{\sqrt{r_{\rm orb}^2+R_{\mathrm{sh}}^2}}{1\rm\,au}\right)^{-2}~{\rm g~cm}^{-3}\,,
\end{equation}
where we set $v_{\rm w}=20\,\rm km\,s^{-1}$. 
The numerical value outside the parentheses in the shocked gas density expression equates to a proton number density of \(n_{\mathrm{ds}}\approx5\times10^{9}{\rm cm}^{-3}\). 
The corresponding pressure $p_{\mathrm{ds}}$ in the shocked region
  \begin{equation}
  p_{\mathrm{ds}}\approx \frac3{16} \rho_{\mathrm{ds}}u_{\mathrm{sh}}^2\,,
  \end{equation}
  decelerates the ejecta according to 
  \begin{equation}
  m_{\mathrm{ej}} \frac{d v_{\mathrm{ej}}}{dt} = -4\pi R_{\mathrm{ej}}^2 p_{\mathrm{ds}}\,.
  \end{equation}
  Here \(m_{\mathrm{ej}}\) is the ejecta mass, \(v_{\mathrm{ej}}\) is ejecta speed (initially the ejecta is assumed to move with \(v_{\mathrm{ej},0}\)), and \(R_{\mathrm{ej}}\) is the distance traveled by  the ejecta:
  \begin{equation}
  R_{\mathrm{ej}}(t)=\int\limits_0^t dt' v_{\mathrm{ej}}(t')\,.
  \end{equation}
  The radius of the forward shock $R_{\mathrm{sh}}$ is determined by the condition that the shocked region density is equal to the value given by Eq.~\eqref{eq:rho_down}:
  \begin{equation}
  \rho_{\mathrm{ds}} = \frac{M_\mathrm{ds}}{4\pi/3\left(R_{\mathrm{sh}}^3-R_{\mathrm{ej}}^3\right)}\,,
  \end{equation}
  where the total mass of the shocked region is given by
  \begin{equation}
  M_\mathrm{ds}(t)=4\pi\int\limits_0^{R_{\mathrm{sh}}(t)}dr' r'{}^2 \rho_{\mathrm{up}}(r')\,.
  \end{equation}
We solve the above system of equations using an iterative method and obtain the time dependence of the forward shock speed, shocked region density, and ejecta speed.
The kinematics of the forward shock and ejecta are shown in Fig.~\ref{fig:shock_speed}, including the shock/ejecta speed (Fig.~\ref{fig:shock_speed}A) and the path covered by the shock (Fig.~\ref{fig:shock_speed}B). For the considered geometry the forward shock propagates in a medium with a constant density during the first \(\sim 12\) hours, Fig.~\ref{fig:shock_speed}B. Afterwards, the upstream medium is characterized by a decreasing density, which results in a period of acceleration of the forward shock. The forward shock speed achieves its maximum value after approximately 3 days. The shock subsequently starts to decelerate, as expected from self-similar solutions in a medium with the density decreasing as \(\rho_{\mathrm{up}}\propto r^{-2}\) \cite{1959book...Sedov}.

\begin{figure}
    \centering
    \includegraphics[width = 0.49\linewidth]{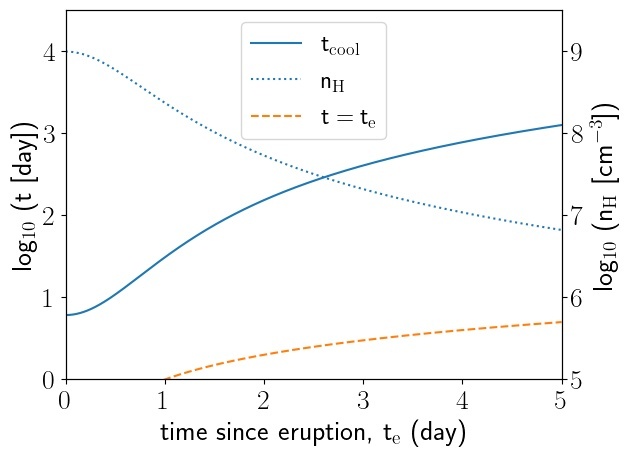}
    \caption{\textbf{Cooling times.} Bremsstrahlung cooling time ($t_{\rm cool}$) of the thermal plasma behind the forward shock as a function of the time since nova eruption ($t_{\rm e}$) in our model (blue solid line), with the time elapsed since eruption also plotted for reference (orange dashed line).  The pre-shock gas density, $n_{\rm H}$, is also shown (blue dotted line, right-hand $y$-axis).}
    \label{fig:tcool}
\end{figure}

For the shock speeds in Figure~\ref{fig:shock_speed}, the post-shock temperature is $>10^8$\,K and the only effective radiative cooling process of the thermal plasma is bremsstrahlung.
Fig.~\ref{fig:tcool} shows that the cooling time of the post-shock gas behind the forward shock (assuming the red giant wind is compressed by a factor of 4) is much longer than the expansion time of the nova for the duration of the H.E.S.S. observations, and so we expect the forward shock to be adiabatic, in contrast to models of slower internal shocks in classical novae \cite{ChomiukReview}.

\begin{figure}[ht]
    \centering
    \includegraphics[width = \linewidth]{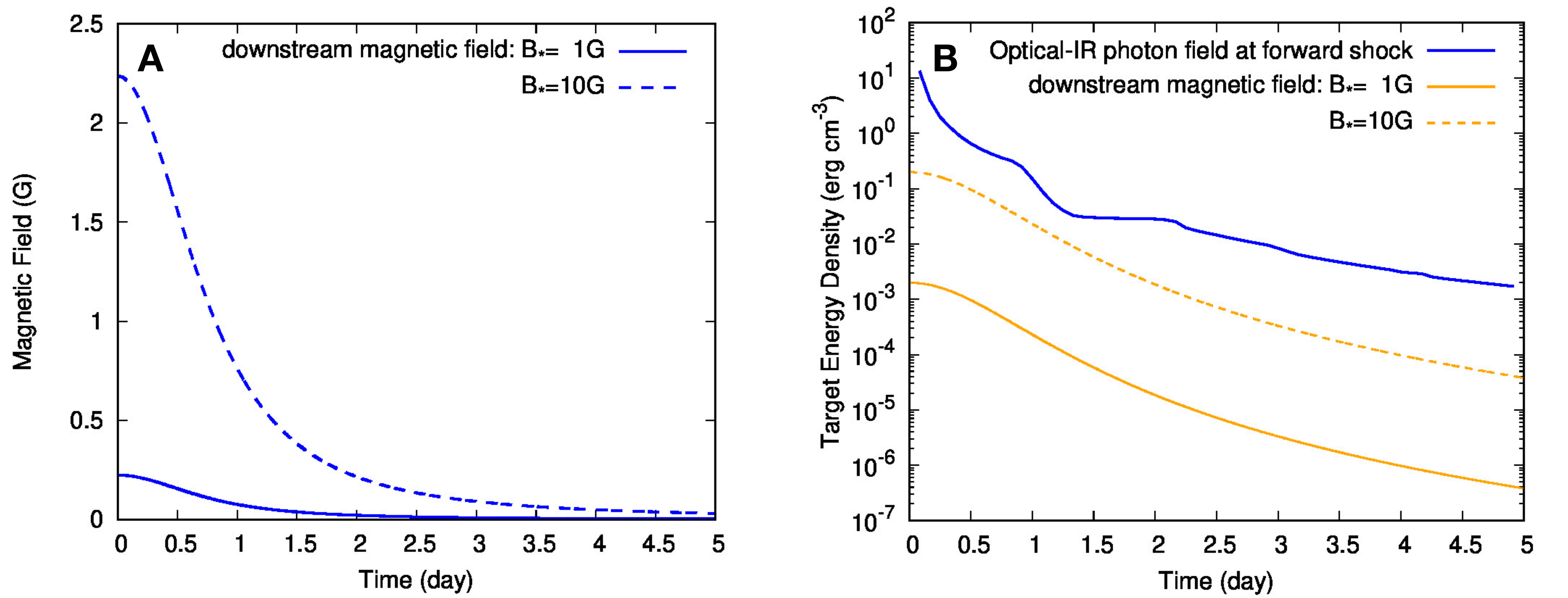}
    \caption{\textbf{Magnetic and photon field evolution.} \textbf{A:} Model magnetic fields as a function of time since the explosion. \textbf{B:} Energy density in the optical photon field at the forward shock and in the magnetic fields as a function of time since the explosion.}
    \label{fig:magnetic_field}
\end{figure}

In the simplest treatment of diffusive shock acceleration at the external shock, assuming Bohm scaling \cite{Drury1983}, the acceleration rate for relativistic particles of charge $q$ is determined by the shock speed and the post-shock mean magnetic field strength, \(B_{\mathrm{sh}}\) as
\begin{equation}\label{eq:acc}
\left.\frac{d E}{dt}\right|_{\mathrm{acc}}=\frac{qB_{\mathrm{sh}}u_{\mathrm{sh}}^2}{\eta c}\,,
\end{equation}
where \(\eta\geq 2\pi\) is a dimensionless parameter 
corresponding to the ratio of the relativistic particle-gyrofrequency to the scattering rate. Its inverse can be taken to characterize the acceleration efficiency and is a measure of the energy in the turbulent magnetic field. If non-linear magnetic field amplification occurs, $\eta$ may fall well below $2 \pi$. This should however be done with care, since magnetic field amplification is unlikely to occur on all scales simultaneously, which is the essential principle of Bohm scaling \cite{2008MNRAS.386..509R}. Particle acceleration to higher energies can be saturated either by escaping the acceleration region (confinement limited) or, in the case of electrons, when inverse Compton (IC)
or synchrotron energy losses overtake the acceleration (cooling limited).
The mean magnetic field strength at the forward shock is determined by the magnetic field of the red giant wind.
We follow the Parker \cite{Parker1958} description of a stellar wind expanding radially outwards from a rotating star, for which the magnetic field near the poles approaches a split-monopole ($B\propto r^{-2}$) and near the equator is swept into a Parker spiral with a toroidal-dominated field ($B\propto r^{-1}$).
Since the nova expansion is understood to be bipolar, and the rotation axis of the red giant should be similar to the orbital axis of the binary system, we adopt the split-monopole approximation for the magnetic field strength at distance $R$ from the red giant:
\begin{equation}
B_{\rm w}(R) = B_* \left(\frac{R}{R_*}\right)^{-2}\,,
\end{equation}
where \(B_*\) is the red giant surface magnetic field, typically of order $1\,\mathrm{G}$ \cite{Kemball1997}.
We consider two cases, adopting 
$B_*=1$ and $10\,{\rm G}$
(baseline and extreme values, respectively). The computed strength of the magnetic field in the blast-wave downstream is shown in Figure~\ref{fig:magnetic_field}A. The wind should be kinetic-energy dominated at a radius of a few times $R_*$, so a limit on the magnetic field (assuming wind expansion at $v_\infty$) is
\begin{equation}
B_{\rm w}(R) < \frac{\sqrt{\dot{M}v_\infty}}{R} \approx 0.2\,\mathrm{G} \left( \frac{R}{1\,\mathrm{au}}\right)^{-1} \;.
\end{equation}
This is consistent with our assumed surface field of order $(1\,\mathrm{G})$ at $R=R_*$. Substantially larger surface fields can be ruled out, as they would imply magnetically dominated winds at large distances.

\paragraph{Maximum Energy:} The VHE H.E.S.S. detection confirms the acceleration of particles to at least several TeV during the ejecta-dominated phase. The Hillas limit \cite{Hillas} for a particle of charge $q=Ze$ in the above radial field (i.e. ignoring strong magnetic field amplification) is
\begin{equation}
E_{\max} = q \frac{u_{\rm sh}}{c}  R_{\rm sh} B(R_{\rm sh}) \approx 10 Z 
\left(\frac{u_{\rm sh}} {5\,000~{\rm km~s^{-1}}}\right)
\left(\frac{B_*}{ 1~{\rm G}}\right)
\left(\frac{R_{\rm sh}}{\rm au}\right)^{-1}~{\rm TeV}
\label{Eq:Hillas1}
\end{equation}
assuming $R_*=0.35$ au. In the first week the shock advances approximately $2-3$ au per day, such that strong magnetic field amplification above that of the background value $B_{\rm w}$ at the forward shock is essential to account for ongoing TeV particle acceleration $>3$ days after the explosion. To facilitate acceleration, the magnetic fields are necessarily amplified upstream of the shock, which can only be triggered by energetic particle currents. For $B_{\rm w}\propto R^{-2}$ the magnetic energy density falls rapidly, suggesting that only a small fraction of cosmic-ray energy needs to be converted to magnetic energy, for example via the Bell instability \cite{Bell2004}, to amplify the fields above background levels. This is an exact analogue of the core-collapse supernova scenario \cite{Bell2013}, for which there is a formalism to predict the maximum energy for a fast shock in a dense wind. We summarize that calculation \cite{Bell2013} below.

Neglecting shock curvature, and assuming isotropy of the particles, the differential accelerating flux (i.e. the rate at which particles increase their energy from $E$ to $E+dE$) at a shock is determined by the shock compression \cite{MalkovDrury}
\begin{equation}
\psi(E) = \frac{\Delta u}{3} E ~N_E
\end{equation}
where $N_E(E)$ is the differential non-thermal particle density, and $\Delta u = u_{\rm sh}-u_{\rm ds}$, i.e. the difference between the upstream and downstream flow velocities. In the model \cite{Bell2013}, particles must generate their own self-confining waves at the expense of an escaping flux. Setting the escaping current density to the accelerating flux of the maximum energy particles when the shock was at a radius $r$ i.e. $j_{\rm esc}(r) = q \psi(E_{\rm max})$, and for simplicity assuming an $N_E \propto E^{-2}$ spectrum, it follows that
\begin{equation}
\label{eq:esc_flux}
\frac{ j_{\rm esc} E_{\rm max}}{q} =  \xi_{\rm esc}\frac{1}{2} \rho_{\rm up} u_{\rm sh}^3 
\end{equation}
where 
\begin{equation}
\xi_{\rm esc} = \frac{\Delta u}{u_{\rm sh}}
\left(\frac{U_{\rm cr}}{\rho_{\rm up} u_{\rm sh}^2/2}\right)
\frac{1}{\log (E_{\rm max}/m c^2)} 
\end{equation}
with $U_{\rm cr}$ the energy density of cosmic-rays above the rest mass energy. The efficiency parameter $\xi_{\rm esc}$ corresponds to the fraction of energy density flux processed by the shock that is lost to the upstream escaping energetic particles, and since $E_{\rm max} \approx 10 TeV$, we expect $\log (E_{\rm max}/m c^2)\approx 10$. 
For acceleration efficiencies of ${U_{\rm cr}}/(\frac{1}{2}\rho_{\rm up} u_{\rm sh}^2)\approx 10\%$, approximately $1\%$ of the energy density flux processed by the shock is lost to the upstream. 
To determine if particles are confined, the fluctuations driven by the non-resonant instability \cite{Bell2004} need sufficient time to grow to a level that can effectively scatter the highest energy particles.
This requires many growth times of the fastest growing instability. The current that passes through a given fluid element at upstream radius $R$ is the integral of the escaping current from the shock over the history of its expansion, diluted by a factor $(r/R)^2$ (here we assume radial magnetic field and ignore deceleration of the shock). Using the definition of the maximum growth rate for cosmic-ray driven instabilities \cite{Bell2004}, and demanding at least $5$ growth times, we find
\begin{equation}
\int \Gamma_{\rm max} dt = \int \sqrt{\frac{\pi}{\rho_{\rm up} c^2}}j_{\rm esc} dt >5
\end{equation}
Using equation \ref{eq:esc_flux} to determine $j_{\rm esc}$, by the time the shock reaches a fluid cell located at radius $R$, the confinement condition is
\begin{equation}
\xi_{\rm esc} q \int_{R_0}^R \sqrt{\frac{\pi}{\rho_{\rm up} c^2}}\frac{\rho_{\rm up} u_{\rm sh}^2}{ E_{\rm max}}{r^2}dr > 5 R^2
\end{equation}
where $R_0$ is the birth location of the shock. 
Differentiating both sides with respect to $R$, we find
\begin{equation}\label{eq:Bell}
E_{\rm max} < \frac{q c \xi_{\rm esc}}{20}  \sqrt{\frac{\dot{M}}{v_{\rm w}}} \left(\frac{u_{\rm sh}}{c}\right)^2\, ,
\end{equation}
which is independent of both the magnetic field strength and shock radius. Eq.~\eqref{eq:Bell} is the so-called confinement limit. This can also be used to estimate the effective scattering magnetic field $B_{\rm eff}$ felt by the maximum energy particles found by equating the above expression for the confinement limit with the first equation in equation \ref{Eq:Hillas1}. This results in a magnetisation parameter of
\begin{equation}
\frac{B_{\rm eff}^2}{4 \pi \rho_{\rm up} u_{\rm sh}^2} > \left(\frac{\xi_{\rm esc}}{20}\right)^2\, .
\end{equation}
This is only a representative value of the upstream field. Stronger magnetic fields concentrated on scales much less than the gyroradius of the maximum energy particles as measured in $B_{\rm eff}$ will not affect their acceleration. Further amplification can also occur in the post-shock medium, though this has only a minor effect on the acceleration rate, as the acceleration time is dominated by the upstream residence time of the diffusing particles. 
Because $E_{\rm max}$ depends only on the properties of the wind, the acceleration time for particles with energy $E_{\rm max}$ is always approximately equal to the instantaneous age of the shock.

Using the shock and wind parameters discussed above for RS Oph, the maximum energy is several TeV. The prediction is consistent with the detection of TeV gamma-rays from RS\,Oph. The confinement limit is the dominant constraint for protons. Electrons must compete also with continuous IC, bremsstrahlung and synchrotron energy losses, so their maximum energy may be limited further. In the numerical leptonic model, the maximum electron energy does not exceed the confinement limit, but this still requires an escaping flux with $\xi_{\rm esc} \approx 0.01$.

\paragraph{Radiative Losses:}
High-energy electrons can interact via different processes losing their energy on time-scales comparable to the acceleration time. The typical cooling times for the synchrotron $t_{\rm syn}$, IC (in the Thomson regime) $t_{\rm IC,Th}$,  and bremsstrahlung $t_{\rm br}$ processes are
  \begin{equation}
  t_{\rm syn}\approx 400 \left(\frac{E}{1\rm\,TeV}\right)^{-1} \left(\frac{B}{1\rm\,G}\right)^{-2}\rm\,s\,;
  \end{equation}
  \begin{equation}
  t_{\rm IC, Th}\approx  15 \left(\frac{E}{1\rm\,TeV}\right)^{-1} \left(\frac{w_{\rm ph}}{1\rm\,erg\,cm^{-3}}\right)^{-1}\rm\,s\,;
  \end{equation}
  and
  \begin{equation}
  t_{\rm br}\approx 10^{6} \left(\frac{n_{\mathrm{ds}}}{10^{9}\rm\,cm^{-3}}\right)^{-1}\rm s\,.
  \end{equation}
  Here \(w_{\rm ph}\) is the energy density of target photons and \(n_{\mathrm{ds}}\) is the number density of particles in the background plasma. These estimates show that for the conditions expected in the production region (see Fig.~\ref{fig:magnetic_field}), the IC mechanism plays the dominant role. For VHE electrons, Klein-Nishina suppression \cite{BlumenthalGould} decreases the rate of IC scattering, and synchrotron cooling might become dominant in this range. We therefore consider these two processes below.

  As the radiative cooling becomes more important with increasing electron energy, we are interested in accurately estimating its contribution for VHE electrons. TeV electrons efficiently up-scatter target photons with energy below the limit set by the Klein-Nishina cutoff, \(<0.25\rm\,eV\).     Thus, optical and infra-red (IR) observations provide information about the intensity of the soft photon fields, which serve as a target for IC losses and emission. Assuming these photon fields result from thermal free-free emission, the expected spectrum in this energy band can be approximated as
  \begin{equation}
  S_\nu\propto \nu^{-2}B_\nu(T(t))\,,
  \end{equation}
  where \(B_\nu\) is the Planck function, \(T\) is the (time-dependent) temperature of the gas producing free-free emission, and the factor \(\nu^{-2}\) appears due to free-free opacity. For the typical post-shock temperatures 
$T>10^8\,{\rm K}$, we are interested in the lower energy part of the spectrum, $h\nu< k_B T$, and the blackbody spectrum can be approximated by the Rayleigh–Jeans law, \(B_\nu\propto \nu^2\). Here $h$ and $k_B$ are Planck and Boltzmann constants, respectively. The relevant part of the spectrum should have a flat spectral energy distribution (SED). 
  In our model calculations we directly use the time-dependent information from the optical B, V, R, and I bands \cite{aavso}, and below the I band we approximate the target photon spectrum by a flat SED component. This approximation agrees with the IR spectra measured from RS Oph after its eruption in 2006 \cite{2007ApJ...671L.157E}.
The time-dependence of the photon target energy density is shown in Figure~\ref{fig:magnetic_field}B.

\begin{figure}[ht]
    \centering
    \includegraphics[width = \textwidth]{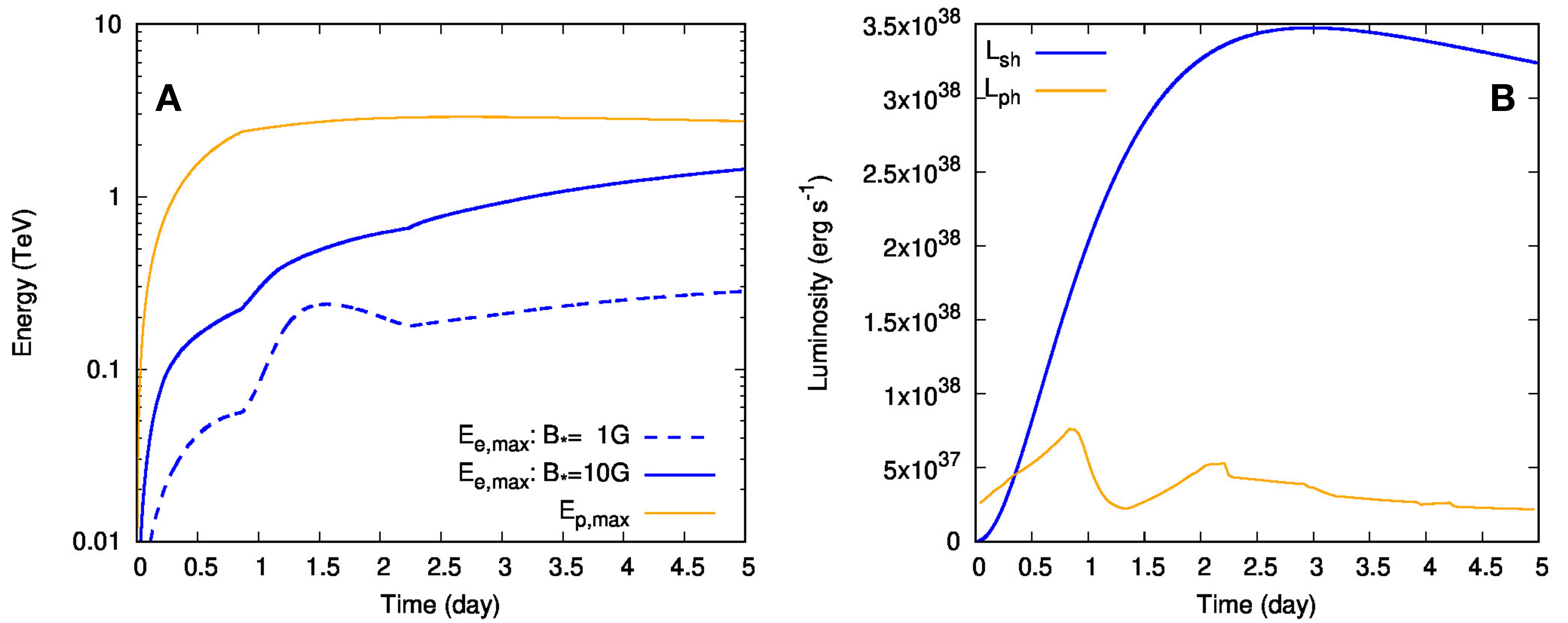}
    \caption{\textbf{Particle energy and luminosity evolution.} \textbf{A:} Maximum energy of accelerated electrons and protons at the forward shock. \textbf{B:} Energy crossing the forward shock per time unit, \(L_{\rm sh}\), and observed optical-IR luminosity of the nova, \(L_{\rm ph}\). The dip occurring around day 1 was in all optical bands.}
    \label{fig:max_energy}
\end{figure}

The evolution of the non-thermal particles follows the acceleration-losses equation: 
\begin{equation}\label{eq:max_energy}
\frac{d E}{dt}=\left.\frac{d E}{dt}\right|_{\mathrm{acc}}+\left.\frac{d E}{dt}\right|_{\mathrm{losses}}\,,
\end{equation}

where the acceleration rate is given by equation \eqref{eq:acc} and  for electrons the loss term accounts for both synchrotron and IC mechanisms, \(\left.\frac{d E}{dt}\right|_{\mathrm{losses}}=\left.\frac{d E}{dt}\right|_{\mathrm{syn}}+\left.\frac{d E}{dt}\right|_{\mathrm{ic}}\).
Relativistic bremsstrahlung is important at early times, but negligible compared to synchrotron and IC by the time the H.E.S.S. observations commenced.
Similarly, for the expected densities, the energy losses for protons are negligibly small.  The maximum energy determined from the solution to equation~\eqref{eq:max_energy} for the two cases considered is shown in Figure~\ref{fig:max_energy}.
For this plot we adopted \(\eta=2\pi\) both for electrons ($\eta_e$) and protons ($\eta_p$). In the case of protons, we assume the maximum energy follows the confinement limit described above with \(\xi_{\rm esc}=10^{-2}\).
For protons the maximum energy does not depend on the magnetic field strength (except for a short time interval after the onset of the acceleration, which is not distinguishable on the scale of the figure). The maximum energy grows rapidly at early time due to the strong magnetic fields close to the star, and the \(\eta=2\pi\) example shown in Figure~\ref{fig:max_energy} likely over-estimates the acceleration rate. Following the rapid growth, the maximum energy saturates at the confinement limit and follows this value for later times. Thus non-linear magnetic field amplification is implicit in the numerical model. The transition from fast to slow cooling is also evident in the electron spectrum .

The obtained maximum energy determines the spectral cut-off of the non-thermal particles, which is injected downstream at each moment:
\begin{equation} \label{eq:ninject}
\frac{d N_{\rm inj}}{dE dt} = \left\{
  \begin{matrix}
    A_{e/p} E^{-\alpha_{e/p}}{\rm exp}{\left[-\left(\frac{E}{E_{\mathrm{max},e/p}}\right)^{\beta_{e/p}}\right]}& E\geq E_{\mathrm{min},e/p}\\
      0 & E<E_{\mathrm{min},e/p}\,.
  \end{matrix}
\right.
\end{equation}
Here \(E_{\mathrm{max},e/p}\) is obtained by solving Eq.~\eqref{eq:max_energy} subject to the limit in Eq.~\eqref{eq:Bell}, for electrons and protons; \(E_{\mathrm{min},e/p}\) is a free parameter set to  \(100\rm\,MeV\) for leptons and \(2\rm\,GeV\) for protons to avoid non-relativistic effects. The normalization coefficient \(A\) is computed by the condition that a fixed fraction, \(\kappa_{e/p}\), of the energy flux through the forward shock, \(L_{\mathrm{sh}}\), is transferred to non-thermal particles:
\begin{equation}
A_{e/p} = \frac{\kappa_{e/p} L_{\mathrm{sh}}}{\int\limits_{E_{\mathrm{min},e/p}}^{\infty}dE E^{1-\alpha_{e/p}}{\rm exp}{\left[-\frac{E}{E_{\mathrm{max},e/p}}\right]}}\,.
\end{equation}
The energy flux through the shock is determined by the upstream medium density and the forward shock speed:
\begin{equation}
L_{\mathrm{sh}} = 2\pi R^2 \rho_{\mathrm{up}}u_{\mathrm{sh}}^3\,.
\end{equation}
The model dependence of this parameter is shown in Figure~\ref{fig:max_energy}B.

The spectrum of non-thermal particles, $n=d N/dE$, is described by the single-zone evolution equation
\begin{equation}
\frac{\partial n}{\partial t} + \frac{\partial }{\partial E}\left[\dot{E}_{\rm losses} n\right] = \frac{d N_{\rm inj}}{dE dt}\,,
\label{eq:S28}
\end{equation}
where the injection term is given by Eq.~\eqref{eq:ninject} and the initial condition is \(n\big|_{t=0}=0\).

\begin{figure}[ht]
    \centering
    \includegraphics[width=\textwidth]{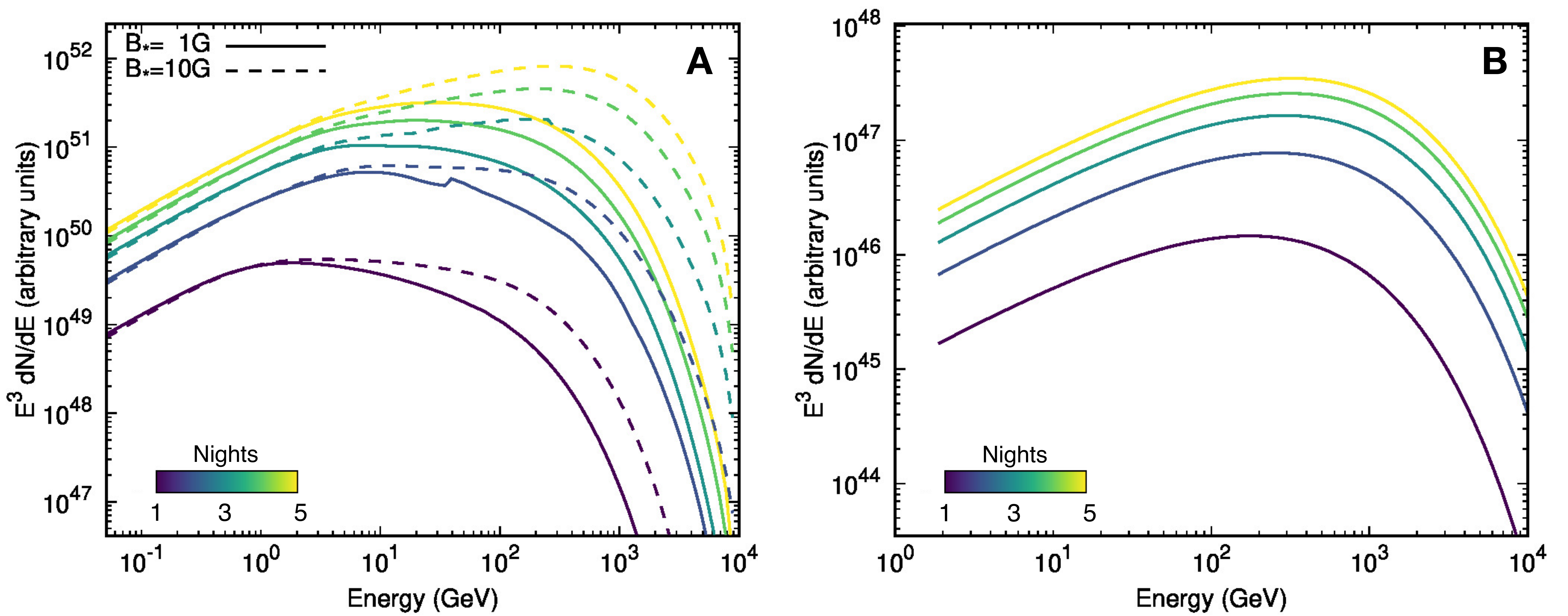}
    \caption{\textbf{Single-zone particle spectra evolution.} \textbf{A:} Electron spectra for five instants in time post-outburst and two different values for the strength of the red giant surface magnetic field. \textbf{B:} Spectra of non-thermal protons for five instants in time (the impact of magnetic field strength is minor, due to confinement limited saturation). }
    \label{fig:particles}
\end{figure}

The non-thermal particle spectra expected in the source at days \(1\), \(2\), \(3\), \(4\), and \(5\) after the
explosion are shown in Figure~\ref{fig:particles}. In the case of non-thermal protons
the time evolution of the spectrum is simple. The impact of cooling on the proton spectrum is minor and we expect an accumulation of non-thermal protons accompanied by minor spectral evolution. In contrast, the
development of the non-thermal electrons is quite complex, and the spectral properties are affected by the
strength of the magnetic field. There are features caused by the decrease in energy loss rates with radius: increasing maximum
energy and spectral hardening caused by the transition from the fast cooling to slow cooling regimes. For the chosen parameters, we do not expect acceleration of electrons to very-high energies during the first day after the explosion for the case of weak external magnetic field. On the other hand, the flux level in the VHE band is smaller than that at GeV energies thus a sufficient VHE flux can be produced by particles with energy exceeding the cutoff energy, in the tail of the distribution. 
Therefore for the radiation calculations we consider the case with $B_*=1\,{\rm G}$, which as discussed above is more consistent with the value expected at the RG surface. 

\begin{figure}[ht]
    \centering
    \includegraphics[width = 1\linewidth]{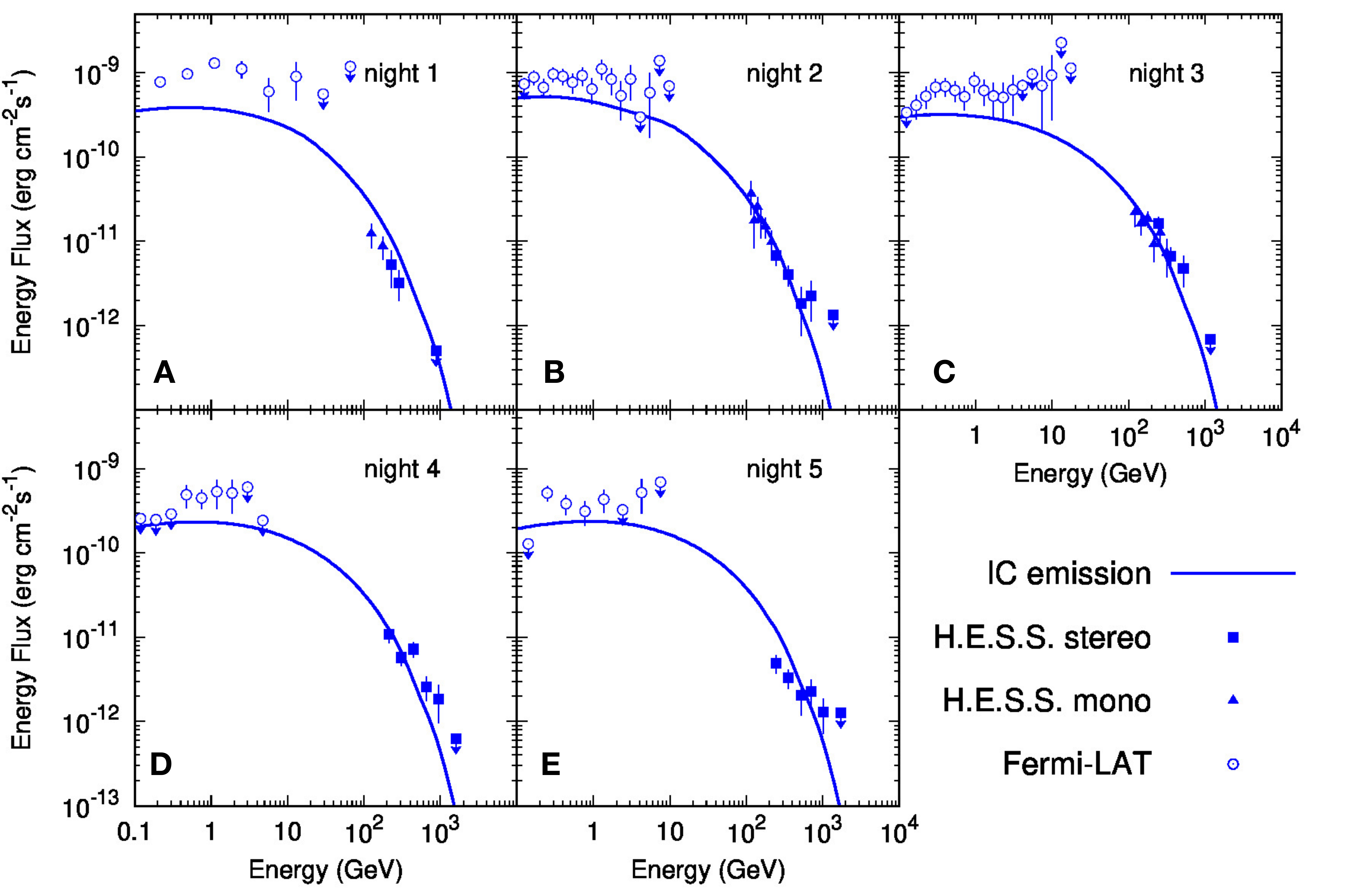}
    \caption{\textbf{Nightly Inverse Compton spectra.} \textbf{A-E} IC emission of shock-accelerated electrons for nights 1-5 respectively, together with spectral measurements obtained with H.E.S.S. (filled points) and \emph{Fermi}-LAT (hollow points). The model parameters are summarized in Table~\ref{tab:model_parameters}.  }
    \label{fig:ic}
\end{figure}

In Figure~\ref{fig:ic} we compare the model IC spectra produced by the shock-accelerated electrons to observational data obtained with
\emph{Fermi}-LAT in the GeV band and H.E.S.S. data obtained in the VHE regime. We set \(\kappa_{e}=0.1\) and
compute the IC spectra for five different instants in time post-outburst: \(t=1\), \(2\), \(3\), \(4\), and \(5\)~days after the
explosion. 
The IC model under-predicts the flux on night\,1; although an increased IC flux for the first night can be obtained by increasing the density of the wind in the model, this change would also accelerate the evolution, with the wind deceleration commencing between days 1 and 2 post explosion. A denser wind is therefore problematic due to the rapid deceleration of the shock (incompatible with multi-wavelength observations) and the need to introduce a suppression of the p-p channel to maintain dominant IC (inconsistent with the theory of shock acceleration).

\begin{figure}[ht]
    \centering
        \includegraphics[width = 1\linewidth]{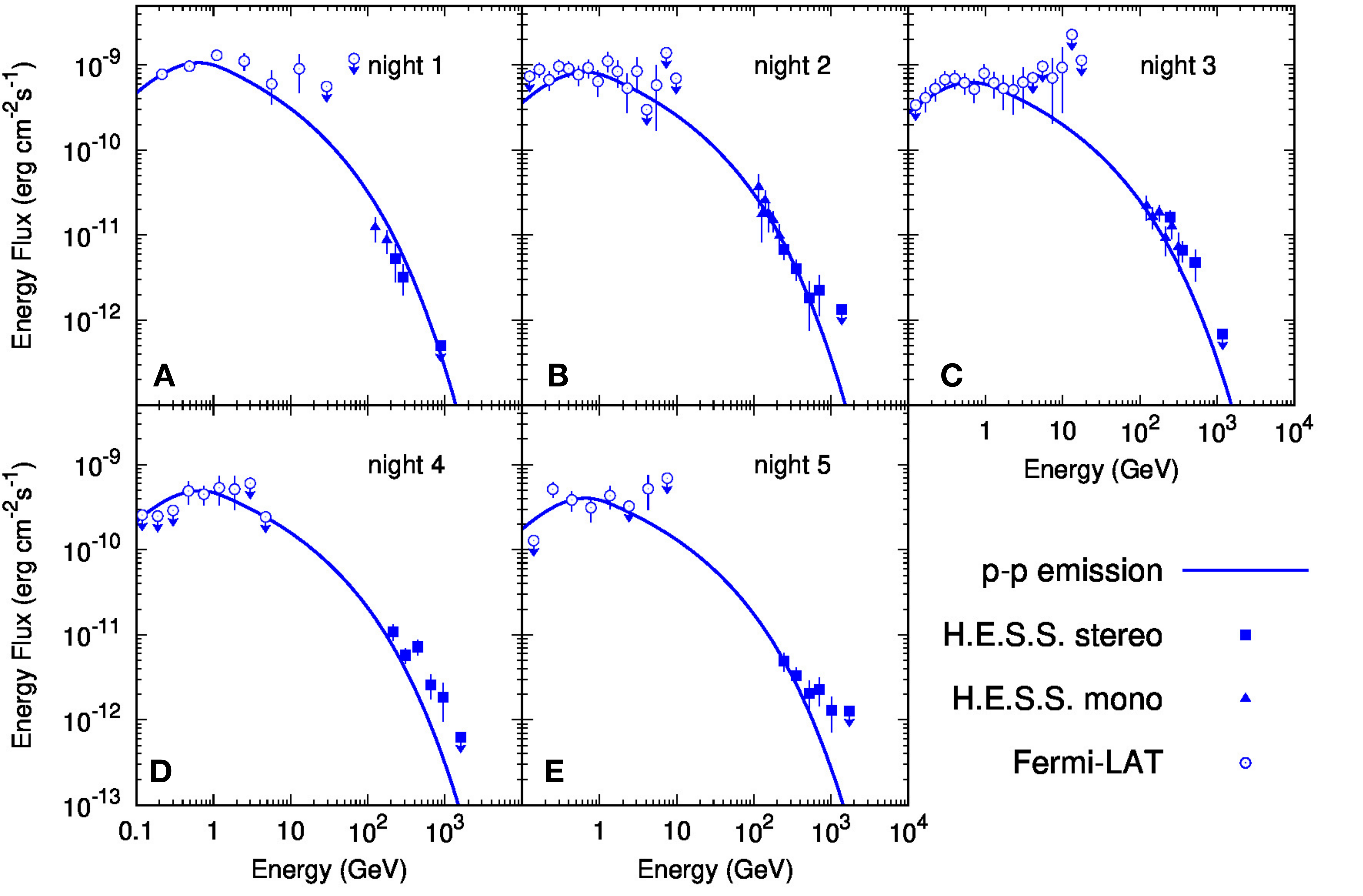}
    \caption{\textbf{Nightly hadronic emission spectra.} Same as Fig. \ref{fig:ic}, but for proton-proton emission. The model parameters are summarized in Table~\ref{tab:model_parameters}.}
    \label{fig:pp}
\end{figure}

In Figure~\ref{fig:pp} we show the result of hadronic emission spectra \cite{Kafexhiu2014} due to collisions between the non-thermal protons and shock compressed external medium for the same times, using  \(\kappa_{p}=0.5\). Here we assume that the gamma-ray emission is produced in a region close to the external shock where the density is twice higher than the averaged value (i.e., the filling factor is \(50\%\) ).  
The inverse of the acceleration-rate efficiency for electrons was set to $\eta_e=10\pi$ and the maximum energy of protons was calculated for $\xi_{\rm esc}=10^{-2}$ and $\eta_{p}=30\pi$.

The local intense photon fields, especially during the first days after the explosion, require consideration of $\gamma-\gamma$ attenuation. We provide some general estimates, which allow us to define the photon frequencies and epochs when $\gamma-\gamma$ absorption might be important. For a source with luminosity \(L\) and size \(R_{\rm sh}\) we can put the following upper limit for the optical depth, \(\tau\), for photons with energy \(\varepsilon\):
  \begin{equation}\label{eq:tau_max}
  \tau\ll\tau_{\rm max}=\frac{0.2\sigma_{\rm T}\varepsilon L}{4\pi R_{\rm sh} c (3.5 m_e^2c^4)}=2 \left(\frac{L}{10^{37} {\rm \,erg\,s}^{-1}}\right)
  \left(\frac{\varepsilon}{1{\rm \,TeV}}\right)
  \left(\frac{R_{\rm sh}}{1{\rm\,au}}\right)^{-1}\,.
  \end{equation}
  Here \(0.2\sigma_{\rm T}\) is the maximum cross-section for the pair creation process expressed in the Thomson cross-section value (\(\sigma_{\rm T}\)). This maximum is achieved for \(\varepsilon h\nu_{\rm target}\approx3.5m_e^2c^4\), where \(h\nu_{\rm target}\) is the target photon energy. The attenuation of TeV gamma-rays occurs due to the interaction with optical~--~IR photons, and  X-rays provide the dominant target for GeV photons. 
  
  Equation~\eqref{eq:tau_max} shows that the $\gamma-\gamma$ attenuation is the most relevant for VHE photons. For this gamma-ray energy range,  optical and IR photons provides the dominant target, and below we compute the corresponding optical target photons. During the first several days after the eruption the typical optical~--~IR luminosity remains at the level of \((1-5)\times10^{37}\rm erg\,s^{-1}\) (see Fig.~\ref{fig:max_energy} and \cite{aavso}), while the typical X-ray luminosity is fainter. For example, X-ray fluxes at the level of \(1 \rm\,keV \, cm^{-2}\,s^{-1}\) were reported at days $3-6$ after the eruption in 2006 \cite{2006Natur.442..276S}.  For the adopted source distance of \(1.4\rm\,kpc\) this corresponds to the X-ray luminosity of \(\sim3\times10^{35}\rm erg\,s^{-1}\). This implies that the $\gamma-\gamma$ attenuation might be important for GeV photons only during the first hours after the eruption, when the source size is small, $R_{\rm sh}\ll 1{\rm\,au}$.

\begin{figure}[ht]
    \centering
    \includegraphics[width = \textwidth]{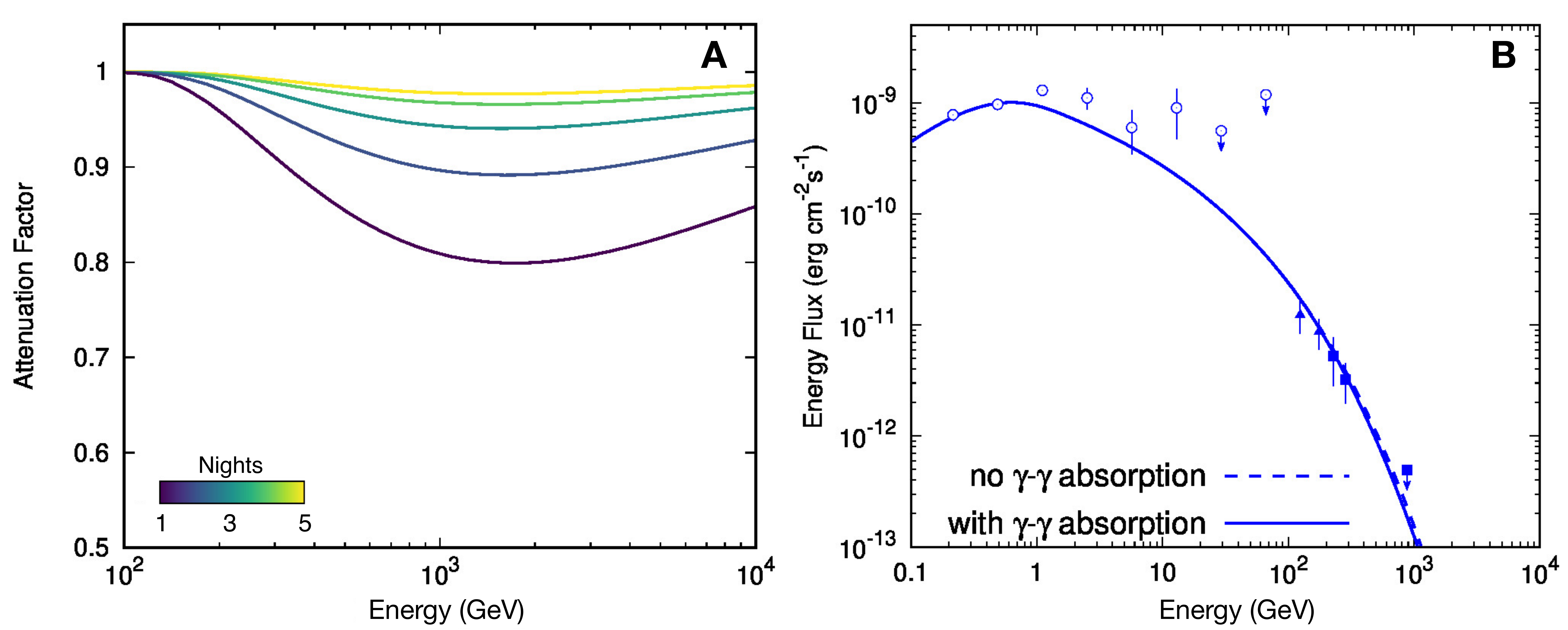}
    \caption{\textbf{Effect of gamma-gamma attenuation.} \textbf{A:} Attenuation factor due to the gamma-gamma absorption on the thermal photons produced by the explosion assuming a distance of $1.4$ kpc. \textbf{B:} The impact of the gamma-gamma absorption on the emission spectrum from night 1 for the source assumed distance of $1.4 {\rm\,kpc}$. The data points are the same as in Fig.~\ref{fig:pp}.}
    \label{fig:gammagamma}
\end{figure}

To estimate its impact we compute also the attenuation factor, \({\rm exp}\left[-\tau\right]\), averaged over the forward shock. The resulting factors, again assuming a distance of 1.4\,kpc are shown in Figure~\ref{fig:gammagamma}  for five different epochs:
\(t=1\), \(2\), \(3\), \(4\), and \(5\)~days after the explosion.  The absorption is not sufficient to account for the observed difference between the HE and VHE bands, and therefore must reflect a feature in the particle population responsible for the gamma rays. The impact of the $\gamma-\gamma$ attenuation is shown in  Fig.~\ref{fig:gammagamma}B where it can be seen that the spectrum transformation by the absorption  
is minor.  We note that the spectral energy distributions in Figs.~\ref{fig:ic} and \ref{fig:pp} account for the attenuation factor obtained for the source distance of $1.4{\rm \,kpc}$ shown in Fig.~\ref{fig:gammagamma}A.

The idealised single zone model presented here does not accurately account for the inhomogeneous structure of the post-shock medium, in which strong magnetic field and gas density gradients naturally develop. 
The purpose of this model is therefore not to precisely fit the data, but rather to reproduce essential features of the system's evolution, characterised by a minimally sufficient number of free variables.

\begin{table}[ht!]
    \centering
    \begin{tabular}{lccc}
     Parameter& Symbol, unit & p-p model & IC model \\ 
      \hline    
    Acceleration slope electrons& \(\alpha_e\)  & -- &\(2.2\)  \\
Acceleration slope protons& \(\alpha_p\)  & \(2.2\) &--  \\
Cutoff exponent electrons& \(\beta_e\)  & -- &\(0.5\)  \\
Cutoff exponent protons& \(\beta_p\)  & \(0.5\) &--  \\
Fraction of energy in electrons&\(\kappa_e\)  & \(0\) &\( 3\%\)  \\
Fraction of energy in protons&\(\kappa_p\)  & \(50\%\) &\(0\)  \\
Acceleration efficiency of electrons&\(\eta_e\)  & -- &\( 10\pi\)  \\
Acceleration efficiency of protons&\(\eta_p\)  & \( 30\pi\) &--  \\
Escape efficiency&\(\xi_{\rm esc}\)  & \(10^{-2}\) &--  \\
Electron low energy cutoff&\(E_{\mathrm{min}}\)&--&\(10^{ 2}m_ec^2\)\\
Proton low energy cutoff&\(E_{\mathrm{min}}\)&\(2.m_pc^2\)&--\\
\hline
RG surface magnetic field& \(B_*,\rm\, G\) & \multicolumn{2}{c}{$1.$} \\
RG radius, au&\(R_{*},\rm\, au\) & \multicolumn{2}{c}{$0.35$}\\
RG mass-loss rate &\(\dot{M}/v_{\rm w}, \rm\,{\rm g~cm}^{-1}\) & \multicolumn{2}{c}{$6.3\times10^{11}$} \\
WD orbit radius&\(r_{\rm orb},\rm\, au\) & \multicolumn{2}{c}{$1.48$} \\
Distance from Earth&\(\rm kpc\) & \multicolumn{2}{c}{$1.4$}\\
\hline
Ejecta initial speed & \(v_{\mathrm{ej},0},~ {\rm km\,s}^{-1}\) &\multicolumn{2}{c}{$3000$} \\
Ejecta mass & \(m_{\mathrm{ej}}\) & \multicolumn{2}{c}{$10^{-7}M_\odot$} \\

    \end{tabular}
    \caption{{\bf Numerical model parameters.} Assumed parameters of the model used in equations \eqref{eq:S1} through \eqref{eq:S28} above, for the two cases of proton-proton (p-p) emission (see Figure~\ref{fig:pp}) and a leptonic Inverse Compton (IC) model (see Figure~\ref{fig:ic}) respectively.}
    \label{tab:model_parameters}
  \end{table}

%TC:endignore
\end{document}